\documentclass{aa}

\usepackage{graphicx} 
\usepackage{txfonts}
\usepackage{hyperref}
\usepackage{gensymb}
\usepackage{bm}
\usepackage{xcolor}

\newcommand{\diff}{\mathop{}\!\mathrm{d}}

\def\bv{\bm{v}}
\def\bw{\bm{w}}
\def\bOmega{\bm{\Omega}}

\newcommand{\nside}{N_\mathrm{side}}
\newcommand{\lmax}{\ell_\mathrm{max}}
\newcommand{\rmax}{r_\mathrm{max}}

\newcommand{\Planck}{\textit{Planck}}
\newcommand{\sevem}{\texttt{SEVEM}}
\newcommand{\commander}{\texttt{Commander}}

\newcommand{\slow}{\textnormal{SLOW}}
\newcommand{\coru}{\textit{Coruscant}}
\newcommand{\magn}{\textit{Magneticum}}

\newcommand{\healpix}{\texttt{HEALPix}}

\begin{document}

\title{Revisiting the CMB large-scale anomalies: The impact of the Sunyaev-Zeldovich signal from the Local Universe}
\titlerunning{Revisiting the CMB large-scale anomalies: The impact of the SZ signal from the Local Universe}
   
\author{
Gabriel Jung\inst{1} \thanks{E-mail: gabriel.jung@universite-paris-saclay.fr},
Nabila Aghanim\inst{1},
Jenny G. Sorce\inst{2,1,3},
Benjamin Seidel\inst{4},
Klaus Dolag\inst{4,5},
Marian Douspis\inst{1}
}
\authorrunning{Jung et al.}
\institute{Université Paris-Saclay, CNRS, Institut d’Astrophysique Spatiale, 91405 Orsay, France 
\and
Univ. Lille, CNRS, Centrale Lille, UMR 9189 CRIStAL, 59000 Lille, France
\and
Leibniz-Institut f\"{u}r Astrophysik (AIP), An der Sternwarte 16, 14482 Potsdam, Germany
\and
Universit\"ats-Sternwarte, Fakult\"at f\"ur  Physik, Ludwig-Maximilians Universität, Scheinerstr. 1, 81679 M\"unchen, Germany
\and
Max-Planck-Institut für Astrophysik, Karl-Schwarzschild-Straße 1, 85741 Garching, Germany
}
    
\date{}
 
\abstract{
The full-sky measurements of the cosmic microwave background (CMB) temperature anisotropies by WMAP and \Planck\ have highlighted several unexpected isotropy-breaking features on the largest angular scales. 

We investigate the impact of the local large-scale structure on these anomalies through the thermal and kinetic Sunyaev-Zeldovich effects. We used a constrained hydrodynamical simulation that reproduced the local Universe in a box of $500\,h^{-1}\,$Mpc to construct full-sky maps of the temperature anisotropies produced by these two secondary effects of the CMB, and we discuss their statistical properties on large angular scales. We show the significant role played by the Virgo cluster on these scales, and we compare it to theoretical predictions and random patches of the Universe obtained from the hydrodynamical simulation \magn. 

We explored three of the main CMB large-scale anomalies, that is, the lack of correlation, the quadrupole-octopole alignment, and the hemispherical asymmetry, in the latest \Planck\ data (PR4), where they are detected at a level similar to the previous releases. We also use the simulated secondaries from the local Universe to verify that their impact is negligible.
} 

\maketitle

\section{Introduction}
\label{sec:introduction}

High-precision observations of cosmic microwave background (CMB) anisotropies in the past two decades have highlighted over a wide range of scales the Gaussian and isotropic nature of the early Universe that corresponds to a standard $\Lambda$CDM Universe with a period of inflation in the simplest models. However, on the largest angular scales, an ensemble of features that break this statistical isotropy or Gaussianity \citep[see e.g.,][and references therein]{Schwarz:2015cma, Abdalla:2022yfr, Aluri:2022hzs} have been observed in the CMB temperature field of both WMAP\footnote{Wilkinson Microwave Anisotropy Probe} and \Planck\ \citep{Planck:2013lks, Planck:2015igc, Planck:2019evm}. This includes the lack of correlation anomaly \citep{Hinshaw:1996ut, WMAP:2003ivt, WMAP:2003elm, Copi:2006tu, Copi:2008hw, Gruppuso:2013dba, Copi:2013cya}, also formulated as a lack of power \citep{Monteserin:2007fv, Cruz:2010ud, Gruppuso:2013xba, Billi:2019vvg, Natale:2019dqm, Billi:2023liq}, the hemispherical asymmetry \citep{Eriksen:2003db, Eriksen:2007pc, Hansen:2008ym, Hoftuft:2009rq, Akrami:2014eta, Adhikari:2014mua, Gimeno-Amo:2023jgv, Kester:2023qmm}, the quadrupole-octopole alignment \citep{deOliveira-Costa:2003utu, Ralston:2003pf, Schwarz:2004gk, Land:2005ad, Gordon:2005ai, Gruppuso:2010up, Copi:2013jna, Marcos-Caballero:2019jqj, Patel:2024oyj}, and the cold spot \citep{2004ApJ...609...22V, 2005MNRAS.356...29C, 2006MNRAS.369...57C, 2010AdAst2010E..77V, Marcos-Caballero:2017iae, Lambas:2023gzy}. The occurrence of one of these so-called large-scale anomalies in CMB simulations of standard $\Lambda$CDM universes is rare, typically below the percent level, and it is even much lower when considered jointly because they mostly appear to be statistically independent \citep{Muir:2018hjv, Jones:2023ncn}. Moreover, future observations of these large cosmological scales will improve the statistical significance of these anomalies, for example, in the CMB polarization field \citep[see e.g.,][including some analyses of the \Planck\ polarization data]{Dvorkin:2007jp, Mukherjee:2015wra, Billi:2019vvg, Chiocchetta:2020ylv, Shi:2022hxc} with the satellite LiteBIRD\footnote{Lite (Light) satellite for the studies of B-mode polarization and Inflation from cosmic background Radiation Detection)} \citep{2018JLTP..193.1048S}, or in the Cosmological Gravitational Wave Background \citep{Galloni:2022rgg, Galloni:2023pie} with LISA\footnote{Laser Interferometer Space Antenna} \citep{2017arXiv170200786A, 2020GReGr..52...81B} \citep{2017arXiv170200786A, 2020GReGr..52...81B}, and they will help us to determine whether they are signatures of unknown physics.

Many of these primordial signatures have been studied. Recent works include scale-dependent primordial non-Gaussianity \citep{Schmidt:2012ky, Byrnes:2015asa, Ashoorioon:2015pia, Adhikari:2015yya, Byrnes:2015dub,Adhikari:2018osh, Hansen:2018pgg}, direct-sum inflation \citep{Kumar:2022zff, Gaztanaga:2024vtr}, cosmic bounces \citep{Agullo:2020fbw, Agullo:2021oqk}, cosmic strings \citep{Jazayeri:2017szw, Yang:2017wzh}, or loop quantum cosmology \citep{Agullo:2021oqk}. However, before any detection of a primordial signature can be claimed, it is necessary to fully characterize the different foregrounds that can leave an imprint on the large-scale temperature fluctuations in the CMB. At the level of the local large-scale structure (LSS), the presence of an unknown foreground has been claimed by \cite{Luparello:2022kqb} based on a cross correlation of \Planck\ and $2$MRS\footnote{2MASS Redshift Survey} data \citep[see however][for a discussion about the significance of this detection]{Addison:2024azr}. Around large spiral galaxies of the nearby Universe ($z<0.02$), the CMB temperature decreases significantly. Moreover, the overall corresponding signal reconstructed empirically seems to reproduce the low multipole behavior of the CMB map \citep{Hansen:2023gra} and might even account for the cold spot \citep{Lambas:2023gzy}. More generally, the local LSS has been shown to contain its own features that deviate from the cosmic mean, with a pancake shape structure with a radius of $\sim 100$ Mpc \citep{2021A&A...651A..15B}, an asymmetry between the northern and southern galactic hemispheres in stellar mass density \citep{Karachentsev:2018ysz}, and an underdensity of up to $200$ Mpc \citep{Whitbourn:2013mwa}, except in our very close neighborhood ($\sim 20$ Mpc) because of the Virgo cluster. 

It is very challenging to understand the extent to which these different anisotropic features of the local Universe can impact the CMB fluctuations. A powerful tool for approaching this issue are constrained cosmological simulations, which can accurately reconstruct the local LSS and its physical properties. An extensive body of literature is devoted to generating these constrained simulations with advanced N-body and hydrodynamical descriptions \citep[see][for a review]{Vogelsberger:2019ynw} and initial conditions reconstructed from galaxy observations \citep[see e.g.,][]{1989ApJ...336L...5B, Kravtsov:2001ac, 2008MNRAS.389..497K, 2010MNRAS.406.1007L, 2013MNRAS.432..894J, 2013ApJ...772...63W, 2013MNRAS.435.2065H, Sorce:2013zha, Sorce:2015yna, McAlpine:2022art}. The objective is to obtain a complete description of the distribution of matter and its properties, consistent with the known local LSS. The recent constrained hydrodynamical simulation called \slow, generated from the initial conditions of \cite{Sorce:2018ekv} with its large volume ($500\,h^{-1}\,$Mpc) and realistic baryonic physics as described in \cite{Dolag:2023xds}, enables  detailed analyses of its different components, such as the synchrotron emission \citep{Boss:2023imc} and the galaxy cluster properties \citep{Hernandez-Martinez:2024wxx}. 

In this work, we focus on the CMB secondary anisotropies related to the Sunayev-Zeldovich (SZ) effect \citep{Zeldovich:1969ff, Sunyaev:1972eq, Sunyaev:1980vz, Sunyaev:1980nv} from the local Universe. We produce high-resolution full-sky maps of its two components, the thermal (tSZ) and kinetic (kSZ) component, derived from the distribution of hot baryonic gas in the \slow\ simulation. We characterize these maps first by measuring their power spectrum, which we compare to predictions from the standard halo model and to the outputs from previous hydrodynamical simulations, namely \magn\ \citep{Dolag:2015dta} and \coru\ \cite{Dolag:2005ay}. We evaluate the role of the nearby Virgo cluster with particular care. On large angular scales, both the kSZ and tSZ signals are significantly dominated by the structures from the very local Universe, meaning that they can carry some anisotropic large-scale features of the local LSS. We investigate their correlation with several well-known CMB large-scale anomalies, namely the lack of correlation, the quadrupole-octopole alignment, and the hemispherical asymmetry. Moreover, we apply the analysis pipeline we developed to study these anomalies to the latest \Planck\ data release (PR4) \citep{Planck:2020olo} to confirm their presence at the levels reported in previous releases. This work is complementary to the analyses of \citep{Gimeno-Amo:2023jgv} and \citep{Billi:2023liq}, and it is based on the same dataset.

This paper is organized as follows. In Sect.~\ref{sec:data} we describe our simulated maps of the thermal and kinetic SZ effects from the local Universe. In Sect.~\ref{sec:planck-analysis} we recall different estimators of several CMB large-scale anomalies, and we apply them to the latest \Planck\ data (PR4). In Sect.~\ref{sec:sz-analysis} we characterize the statistical properties of the local tSZ and kSZ maps by studying their power spectra for the first time, and we examine the extent to which they can contribute to the CMB large-scale anomalies. Finally, we draw our conclusions in Sect.~\ref{sec:conclusion}.

\section{Data and simulations}
\label{sec:data}

\subsection{Hydrodynamical simulations}
\label{sec:simulations}

To construct the tSZ and kSZ signals from the local Universe, we used several hydrodynamical simulations that we present in this section. Our baseline simulation was \slow, which is a recent and large constrained hydrodynamical simulation. We compared it to two other simulations that were studied in detail in the literature: the smaller constrained simulation \coru, and the unconstrained \magn. We compare our results with these simulations to verify the impact of improvements in reconstructing the local LSS, and to determine how it differs from an average patch of our $\Lambda$CDM Universe.

\subsubsection{The \slow\ constrained simulation}
\label{sec:slow}

\slow\ is a hydrodynamical simulation that reproduces the CMB large-scale anomalies of the local Universe in a box of $500\,h^{-1}\,$Mpc. It contains $2\times1536^3$ dark matter and gas particles and assumes a \Planck-like cosmology (see Table~\ref{tab:slow}). For a detailed description of its production, we refer to \citet{Dolag:2023xds, Hernandez-Martinez:2024wxx} and references therein. We summarize the most important steps below.

The key ingredient of constrained simulations is the initial conditions, which are reconstructed from observations of the local Universe using different methods. For \slow, the initial conditions are derived following the approach described in \cite{Sorce:2018ekv}. They are based on the use of galaxy peculiar velocities that were derived from the CosmicFlows-2 data \citep{Tully:2013wqa} to trace the full matter distribution. Larger and more recent datasets such as CosmicFlows-3 and 4 \citep{Tully:2016ppz, Tully:2022rbj}, which include CosmicFlows-2, have been produced. While an anomalous bulk flow was recently reported both by \cite{Whitford:2023oww} and \cite{Watkins:2023rll} in CosmicFlows-4, we stress that all data are fully compatible with each other and with the CosmicFlows-2 mean bulk flow value, as shown in figure 14 of \citep{Whitford:2023oww}. As a consequence, we also expect the initial conditions derived from CosmicFlows-2, -3, and -4 to agree with each other.

After the initial conditions were reconstructed from the galaxy peculiar velocities, they were evolved using the code \textsc{OpenGadget3} \citep{2023MNRAS.526..616G}, which was developed from the standard TreeParticleMesh code \textsc{Gadget3} \citep{Springel:2005mi} with an improved smooth particle hydrodynamics solver \citep{2016MNRAS.455.2110B}. In addition to the standard gravitational evolution, the physics of baryonic matter (e.g., cooling, star formation, winds, and feedback from active galactic nuclei) is included through advanced subgrid models that were previously validated with the \magn\ hydrodynamical simulation; see Sect.~\ref{sec:magneticum}.

\begingroup
\setlength{\tabcolsep}{5pt} 
\renewcommand{\arraystretch}{1.5} 
\begin{table}
\caption{Cosmological parameters of the \slow\ simulation}
    \begin{center}
      \begin{tabular}{|cccccc|}
        \hline
        $\Omega_\Lambda$ & $\Omega_\mathrm{m}$ & $\Omega_\mathrm{m}$ & $\sigma_8$ & $n_s$ & $h$ \\
        \hline
         $0.692885$  & $0.307115$ & $0.0480217$ & $0.829$ & $0.961$ & $0.6777$ \\
        \hline
        \end{tabular}
      \end{center}
\tablefoot{The cosmological parameters of the \slow\-constrained simulation are compatible with \Planck\ observations \citep{Planck:2013pxb}.}

    \label{tab:slow}
\end{table}
\endgroup

Several constrained numerical simulations that used the initial conditions reconstructed from the CosmicFlows-2 dataset were produced, and comparisons with observations conducted at different scales showed that they reproduced the properties of the local Universe. At the scale of our galaxy, \citep[e.g.,][]{Carlesi:2016qqp} showed that the local large structure induces the pair formation of the Milky Way  and Andromeda. At the scale of galaxy clusters, replicas of nearby clusters such as Virgo, Centaurus, Hydra, and Perseus were studied in detail by \cite{Sorce:2015yna, Sorce:2018skr, Sorce:2018ekv, Sorce:2021mzn, Lebeau:2023klb}. At larger scales, the filaments connected to the Coma cluster were recovered \citep{Malavasi:2023yrx}, and velocity waves associated with the massive clusters of our local environment were shown to agree with data \citep{Sorce:2023ucz}. \\
The baseline-constrained simulation we used, \slow, was also tested against observations. A large number of the most massive observed galaxy clusters were cross identified in \slow and were found to be close to their actual position in the sky. The masses were within the observational uncertainties \citep{Hernandez-Martinez:2024wxx}. On larger scales, the observed overdensity of massive galaxy clusters up to $100$ Mpc and the underdensity of galaxy clusters in the same sphere, as well as the asymmetries between southern and northern directions, are also well reproduced in the \slow\ simulation \citep{Dolag:2023xds}.

\subsubsection{The \coru\ constrained simulation}
\label{sec:coruscant}

We considered another constrained simulation, called \coru, which focuses on the very local Universe. It describes the local structures in a sphere with a radius of $\sim 110$ Mpc that is embedded in a box of size $\sim 240\,h^{-1}\,$Mpc and contains about $108$ million particles in total (gas and dark matter). The initial conditions were reconstructed from the observed galaxy density field in the IRAS\footnote{Infrared Astronomical Satellite} $1.2$-Jy survey \citep{1995ApJS..100...69F}.

A first version of the simulation was described in detail in \cite{Dolag:2005ay} \citep[see also ][for an earlier work based on similar initial conditions in the dark matter only case]{2002MNRAS.333..739M}, and was later updated to include an earlier version of all the ingredients also used the \magn\ simulation (for other works that analyzed this most recent version of the \coru\ simulation, see, e.g., \cite{Planck:2015grh, Dolag:2015dta, Coulton:2021ekh}).

\subsubsection{The \magn\ simulation}
\label{sec:magneticum}

The final hydrodynamical simulation we studied is \textit{Box2} of the \magn\ simulation\footnote{\url{http://www.magneticum.org/simulations.html}}. Its unconstrained initial conditions consist of $2\times1584^3$ dark matter and gas particles in a box with a size of $352\,h^{-1}\,$Mpc, and the run includes the ingredients of baryonic physics detailed in \cite{Hirschmann:2013qfl, Dolag:2015dta}. 

\subsection{tSZ and kSZ maps of the local Universe}
\label{sec:sz-maps}

The CMB large-scale anomalies of the Universe leaves imprints on the CMB anisotropies through different effects \citep[for a review, see e.g.,][]{Aghanim:2007bt}. One of the most important imprints is the thermal SZ effect \citep{Zeldovich:1969ff, Sunyaev:1972eq}. It consists of the inverse-Compton scattering of CMB photons by electrons in the ionized gas and is mainly located in galaxy clusters. It is noted tSZ. The tSZ effect is proportional to the integral of the electron gas pressure along the line of sight. This corresponds to a spectral distortion of the CMB temperature fluctuations, which can be written as
\begin{equation}
    \label{eq:tSZ}
    \frac{\Delta T^\mathrm{tSZ}}{T_\mathrm{CMB}}(\nu, \bm{\Omega}) = g(\nu) y(\bm{\Omega}),
\end{equation}
where $g(\nu)$ is an overall factor depending on the frequency $\nu$, and the Compton-$y$ parameter describes the signal over the sky. These terms are given by 
\begin{equation}
    \label{eq:tSZ-details}
\begin{split}
    g(\nu)&\equiv x\,
    \mathrm{coth}\left(x/2\right)-4
    \quad\text{with}~x\equiv\frac{h\nu}{k_\mathrm{B}T_\mathrm{CMB}},\\
    y(\bm{\Omega}) &\equiv \frac{\sigma_T}{m_e c^2} \int_0^{r_\mathrm{CMB}} \diff r\, P_e(\bm{\Omega}, r),   
\end{split}
\end{equation}
where $h$ is the Planck constant, $k_B$ is the Boltzmann constant, $T_\mathrm{CMB}$ is the CMB temperature, $c$ is the speed of light, and $\sigma_T$ is the scattering cross section. The electron gas pressure is $P_e=k_B n_e T_e$, where $T_e$, $n_e$, and $m_e$ are the electron temperature, density, and mass, respectively.

The kinetic SZ (kSZ) effect \citep{Sunyaev:1980vz, Sunyaev:1980nv} is induced by the bulk motion of the electron gas that moves with a peculiar velocity $\bm{v}_e$ in the direction of the line of sight. It can be written as
\begin{equation}
    \label{eq:kSZ}
    \frac{\Delta T^\mathrm{kSZ}}{T_\mathrm{CMB}}(\bm{\Omega}) =
    -\frac{\sigma_T}{c}\int_0^{r_\mathrm{CMB}} \diff r\, \bm{\Omega} \cdot \bm{v}_e(\bm{\Omega}, r).
\end{equation}
The kSZ effect has the same blackbody dependence as the CMB primary fluctuations. In amplitude, it is dominated by its tSZ counterpart, except when the tSZ effect is null at a frequency of about 220~GHz. 

Using the numerical code \textsc{SMAC}\footnote{\url{https://www.mpa-garching.mpg.de/~kdolag/Smac/}} \citep{Dolag:2005ay}, we can directly evaluate the integrands in Eqs.~\eqref{eq:tSZ-details},~\eqref{eq:kSZ} throughout the \slow\-constrained hydrodynamical simulation for a \healpix\ tessellation scheme. This allows us to perform the integral in a range that is limited by the simulation volume. It is also possible to push the integration further, for example, to compute the full SZ signal. This requires generating light cones from the simulation \citep[see][for an example based on the \magn\ simulation]{Dolag:2015dta}. However, we are only interested in the SZ signal from the local Universe, and we restricted the computations to the volume of the \slow\ simulation. 

\begin{figure*}
    \centering
    \includegraphics[width=0.49\linewidth]{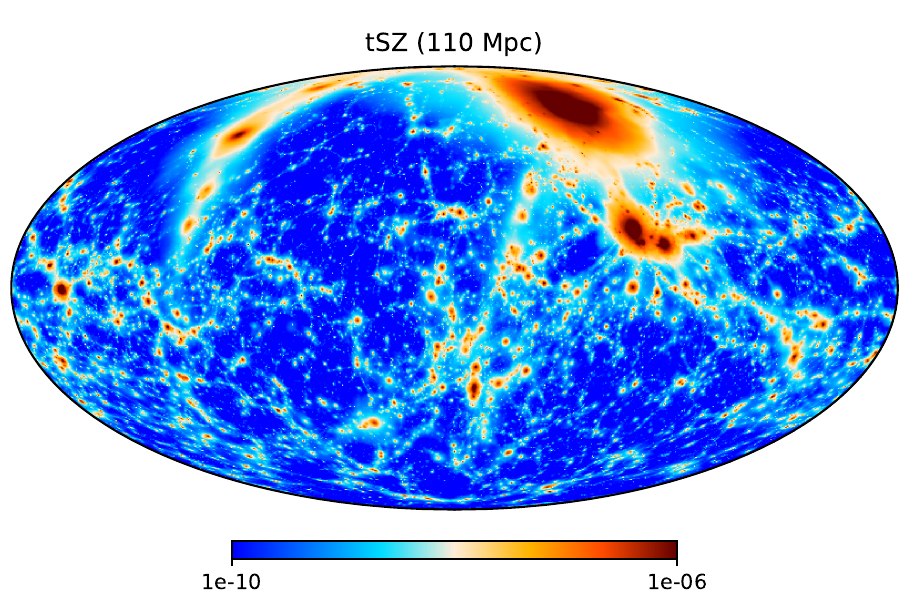}
    \includegraphics[width=0.49\linewidth]{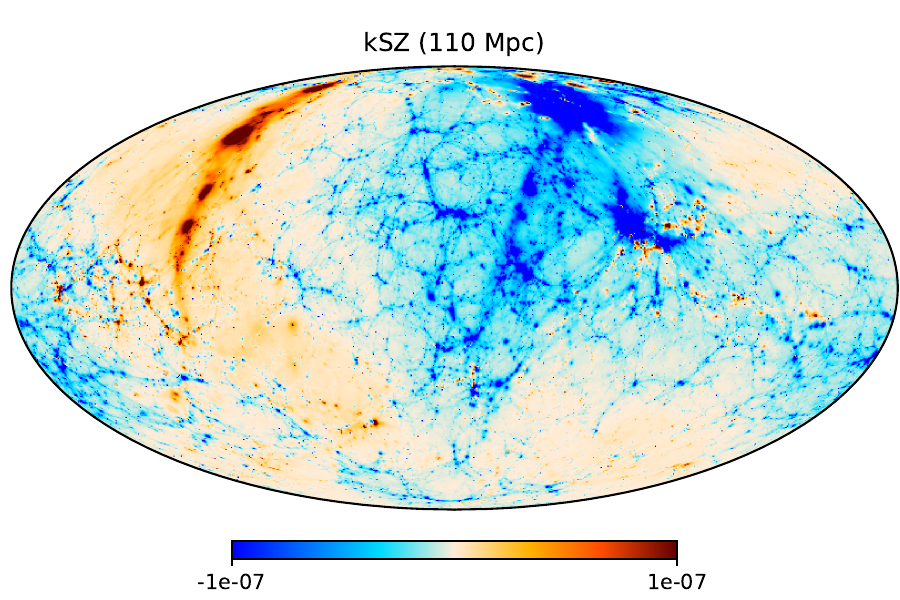}
    \includegraphics[width=0.49\linewidth]{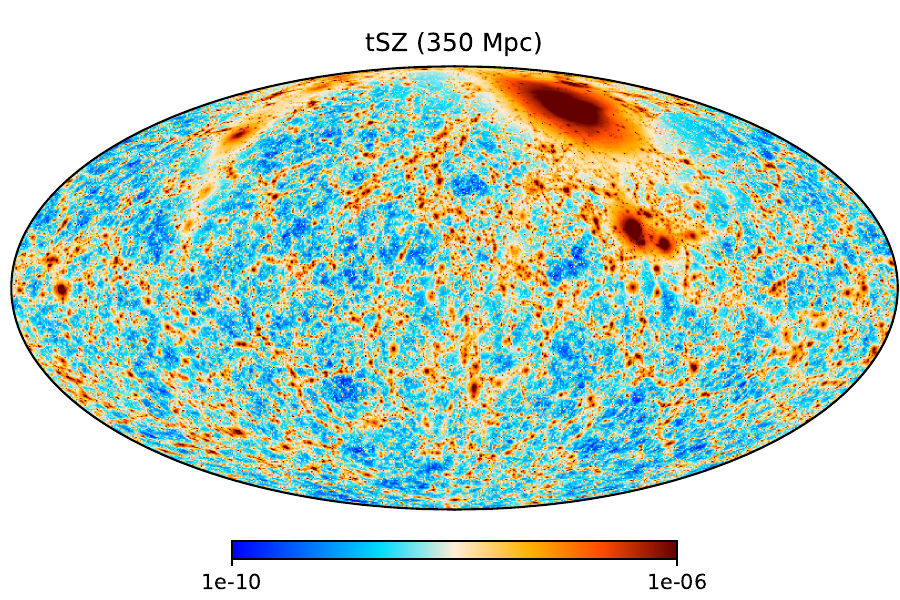}
    \includegraphics[width=0.49\linewidth]{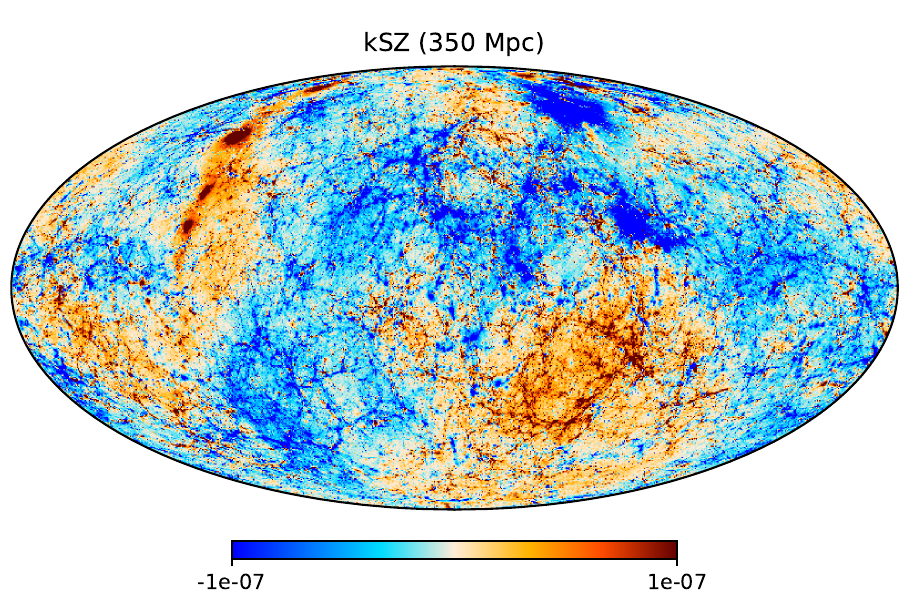}
    \caption{tSZ ($y$-Compton) and kSZ signals from the local Universe in the \slow\ constrained simulation. The upper panels only include the most local structures (up to $110$ Mpc, denoted \slow-110), and the lower panels use almost the full volume of the box (up to $350$ Mpc, denoted \slow-350). The color scale is logarithmic in the tSZ maps and symmetric logarithmic for the kSZ maps. The main characteristics of these maps are recalled in Table~\ref{tab:maps}.}
    \label{fig:maps}  
\end{figure*}

To obtain maps of the local tSZ (Compton-$y$) and kSZ signals, we started the integration from the observer position, which was chosen so that the position of the largest simulated structures on the sky mimicked that of their observed counterparts as much as possible.\footnote{To be exact, we started the integration at $5$ Mpc from the observer, which removed the Milky Way foreground effect as well as all galaxies in the local Group.} By construction of the constrained initial conditions, the observer position is very close to the center of the \slow\ box. Then, to take advantage of the full constrained volume, we performed the integration up to $350$ Mpc around it (the maps we obtained are referred to as \slow-350). We also generated maps focused on the closest structures, up to $110$ Mpc (constrained volume of \coru), from \slow\ (the maps we obtained are referred to as \slow-110) and \coru, both for comparison between the two generations of constrained simulations and for the direct study of close-by structures while avoiding contamination from the background. These maps have a resolution of $\nside=1024$ (pixels with a size of $\sim 3.4$ arcmin) and constitute the core ingredient of this paper. They are shown in Fig.~\ref{fig:maps}. 

On the largest scales, these maps are highly anisotropic as the signal is dominated by a few extended structures from the very local Universe. The most prominent structure, located in the upper right corner of each plot, corresponds to the Virgo galaxy cluster. It leads to a large positive (negative) contribution in the tSZ (kSZ) maps, respectively. Several other well-known galaxy clusters can be cross-identified in this simulation, with a good match of position and mass, as discussed in detail in \cite{Hernandez-Martinez:2024wxx}.

The observed and simulated positions of these clusters in the sky differ slightly, however, as does their distance from the observer. These differences are mainly due to intrinsic limitations of the initial conditions reconstruction method, which causes structures to be slightly displaced with respect to their expected position. These differences in the position of the structures may also arise when the reconstructed initial conditions are based on different datasets, which leads to slightly different values of the bulk flow. To estimate the impact of these differences, we produced a set of maps (\slow-near) in which the observer position was shifted $1$ Mpc in different directions ($14$ new positions per maps) in the \slow\ box. We also tested a larger displacement of $2$ Mpc, which produced very similar conclusions.

Finally, to verify how the local tSZ and kSZ signals compare to random patches of the Universe, we generated sets of maps in which the observer positions were selected in very distant positions (more than $100$ Mpc between each) in both \slow\ (SLOW-grid) and \magn\ ($27$ maps for each). This allowed us to obtain almost independent realizations of the SZ signals.

A list of the different maps and of their main characteristics is given in Table~\ref{tab:maps}. Their statistical properties are discussed in detail in Sect.~\ref{sec:sz-analysis}.

\begingroup
\setlength{\tabcolsep}{6pt} 
\renewcommand{\arraystretch}{1.5} 
\begin{table*}
    \caption{Characteristics of the tSZ and kSZ maps constructed from the \slow, \magn\ and \coru\ hydrodynamical simulations}
    \begin{center}
      \begin{tabular}{|l|l|ccc|}
        \hline
        Name & Description & Number  & $\rmax$ (Mpc) & $\nside$ \\
        \hline
        \textbf{SLOW-110} & Local Universe (from \slow) & $1$ & $110$ & $1024$ \\
        \textbf{SLOW-350} & Local Universe (from \slow)& $1$ & $350$ & $1024$ \\
        \textbf{SLOW-near} & Near local Universe (observer position moved by 1 Mpc) & $14$ & $110$ & $1024$ \\
        \textbf{SLOW-grid} & Grid of observer positions all over the \slow\ simulation & $27$ & $110$ & $1024$ \\
        \hline
        \textbf{MAGNETICUM} & Same as \textbf{SLOW-grid}, for the \magn\ simulation & $27$ & $110$ & $1024$ \\     
        \hline
        \textbf{CORUSCANT} & Local Universe (from \coru) & $1$ & $110$ & $1024$ \\ 
        \hline
        \end{tabular}
      \end{center}
    \label{tab:maps}
\end{table*}
\endgroup

\subsection{Planck CMB data}
\label{sec:planck-data}

For the analyses presented in Sect.~\ref{sec:planck-analysis}, we used the CMB temperature datasets from the latest \Planck\ release \citep[for details about the PR4 release, see][]{Planck:2020olo}. They include foreground-cleaned CMB maps produced by the component separation methods \sevem\ \citep{Fernandez-Cobos:2011mmt} and \commander\ \citep{Eriksen:2007mx}. They are available at the \Planck\ Legacy Archive\footnote{\url{http://pla.esac.esa.int/pla/}} (PLA). We also used the associated simulated CMB maps ($600$ and $100$ for \sevem\ and \commander, respectively), which are accessible from the National Energy Research Scientific Computing Center (NERSC).\footnote{\url{https://portal.nersc.gov/project/cmb/planck2020/}}

The main mask we used is the $2018$ \Planck\ common mask. In addition, to confirm the robustness of our results against astrophysical contamination, we also considered masks covering a larger area near the galactic plane. All these masks can be found in the PLA.

We are primarily interested in large-scale effects and therefore did not work directly with the data at full resolution ($\nside=2048$, and a Gaussian beam with an FWHM of $5'$). We instead used a downgraded resolution ($\nside=64$, and $160'$ Gaussian beam), as in \citet{Planck:2019evm}. A map with this low resolution and large beam washes out small-scale variations of the signal, in particular, the variation that is associated with structures such as clusters of galaxies. This leads to a negligible impact on the overall CMB signal at large scales. After downgrading, we set all pixels with a value below $0.95$ to $0$ for the masks to maintain their binarity.

Before we estimated the different statistics of interest introduced in the following section, the masks were applied to the CMB \Planck\ data maps and their simulated counterparts. We then removed the monopole and dipole of the resulting masked maps. This removed any large-scale coherent signal such as that potentially associated with a large-scale bulk flow. As the mask itself can have a strong impact on the large-scale statistics we measured, it was necessary to fill in the masked regions. We used the diffusive inpainting technique \citep[see for example ][for details]{Gruetjen:2015sta}.

\section{Large-scale anomalies in \Planck\ CMB data}
\label{sec:planck-analysis}

In this section, we investigate several well-studied large-scale anomalies of the CMB temperature anisotropies using the most recent data available from \Planck\ (PR4; see Sect.~\ref{sec:planck-data} for details). For each anomaly, we briefly recall the historical definition and a corresponding standard estimator before we apply it to the \Planck\ observed and simulated CMB maps. 

\subsection{Lack of correlation}
\label{sec:lack-of-correlation}

One surprising feature of the CMB anisotropies that was first observed in \textit{COBE}\footnote{Cosmic Background Explorer} data \citep{Hinshaw:1996ut} is that the two-point angular correlation function $C(\theta)$ (2PACF) seems to vanish on large angular scales ($\theta>60\degree$). To quantify this so-called lack of correlation, the standard choice of statistics is the quantity noted $S_{1/2}$. It is written as
\begin{equation}
    \label{eq:S1/2}
    S_{1/2} \equiv \int_{-1}^{1/2} \left[C(\theta)\right]^2 \diff (\cos\theta).
\end{equation}
It was originally introduced in \citep{WMAP:2003elm} and has been used in many works based on WMAP or \Planck\ data \citep[see for example][]{Copi:2006tu, Copi:2008hw, Gruppuso:2013dba, Copi:2013cya, Schwarz:2015cma, Planck:2013lks, Planck:2015igc, Muir:2018hjv, Planck:2019evm, Jones:2023ncn}. We follow the computation steps of \citet{Copi:2008hw}, who evaluated $S_{1/2}$ directly from (pseudo-) power spectrum measurements. 

To estimate this pseudo-$C_\ell$ from the \Planck\ CMB maps, we used the numerical code \texttt{NaMaster}\footnote{\url{https://github.com/LSSTDESC/NaMaster}} as described in \cite{Alonso:2018jzx} \citep[for earlier works on pseudo-$C_\ell$ estimation, see][]{Wandelt:1998qd, Hivon:2001jp, Hansen:2002zq, Tristram:2004if}. With this method, which was constructed to handle masked maps, we did not need to perform the inpainting step after applying the \Planck\ common mask.

\begin{figure}
    \centering
    \includegraphics[width=0.99\linewidth]{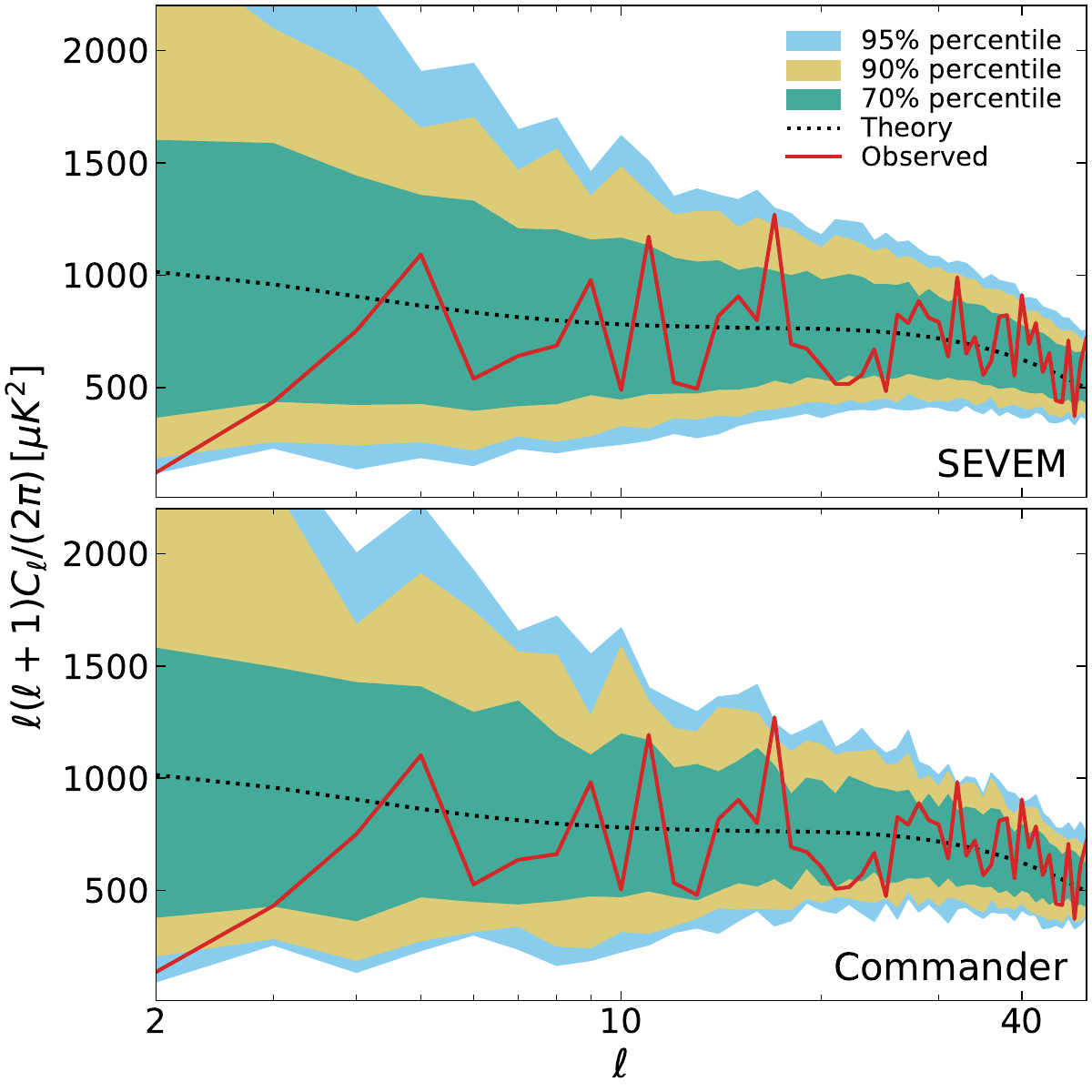}
    \caption{Pseudo-$C_\ell$ of the \Planck\ PR4 CMB temperature maps from \sevem\ (top panel) and \commander\ (bottom panel). The solid red lines are measured from cleaned observed maps, and the dotted black lines are theoretical predictions computed using \texttt{CAMB}. The colored areas are determined from simulations and show their distribution around the median. For the two component separation methods, the observed quadrupole belongs to the lowest $5\%$ of the simulated values (only $15$ \sevem\ simulations out of $600$ have a lower quadrupole, and $4$ for \commander). Only $15\%$ of the simulations have a smaller octopole.}
    \label{fig:pseudo-cl}  
\end{figure}

Our results are shown in Fig.~\ref{fig:pseudo-cl}. As is known since \textit{COBE} observations \citep{1993AdSpR..13l.409B}, which were later confirmed in WMAP data \citep{WMAP:2003elm}, the observed quadrupole has a low amplitude that is one of the lowest in the simulations ($2.5\%$ and $5\%$ of the lowest in \sevem\ and \commander\ simulations, respectively). The octopole moment is also rather low ($15$th percentile). This cannot be considered as anomalous, however, as was shown in \citep{Efstathiou:2003wr}, and it does not explain the lack of correlation effect by itself.

\begin{figure}
    \centering
    \includegraphics[width=0.99\linewidth]{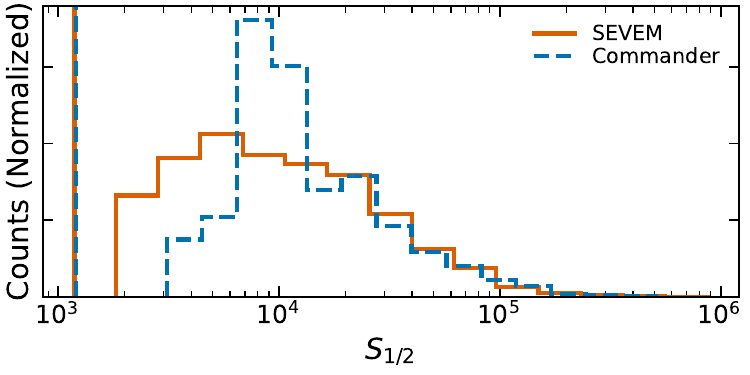}
    \caption{Lack of correlation on large angular scales ($\theta>60\degree$) in the cleaned \Planck\ PR4 CMB temperature maps. We show the distribution of the statistic $S_{1/2}$ (Eq.~\ref{eq:S1/2}) measured in the \Planck\ PR4 simulations, and the vertical lines (on the left) are the observed values. The dashed blue and solid orange lines correspond to the \commander\ and \sevem\ maps, respectively. None of the measured $S_{1/2}$ in the simulations is as low as in the observations.}
    \label{fig:s_half}  
\end{figure}

In Fig.~\ref{fig:s_half}, we show the quantity $S_{1/2}$ computed from the \Planck\ PR4 pseudo-$C_\ell$ of Fig.~\ref{fig:pseudo-cl}. The observed value is clearly lower than in all the simulations. This confirms that the 2PACF is anomalously close to zero on angular scales ($\theta>60\degree$).

\subsection{Alignment of the quadrupole and octopole}
\label{sec:planck-alignment}

Another striking feature of the CMB temperature fluctuations is the alignment of the quadrupole and octopole \citep[first reported in WMAP data][]{deOliveira-Costa:2003utu}, which can be studied using multipole vectors \citep{Maxwell:1865, Copi:2003kt, 2004JPhA...37.9487D}. These multipole vectors constitute an alternative to the standard decomposition of CMB anisotropies into spherical harmonics, where the information contained in each multipole $\ell$ is given by a real constant and $\ell$ unit vectors $\bv_\ell^j$ (with $j=1\cdots\ell$).

To estimate low-$\ell$ multipole vectors in the \Planck\ PR4 maps, we used the optimized numerical code \texttt{polyMV}\footnote{\url{https://oliveirara.github.io/polyMV/}} presented in \citet{Oliveira:2018sef}. We then followed the analyses of \citet{Schwarz:2004gk, Copi:2005ff} where the key ingredient for studying alignments is the area vectors, which are defined as cross-products between multipole vectors at a given $\ell$, 
\begin{equation}
    \label{eq:w-cross}
    \bw_\ell^{(i,j)} \equiv \bv_\ell^i \times \bv_\ell^j.
\end{equation}
There is thus one vector $\bw_2$ for the quadrupole, corresponding to one preferred direction, and three vectors ($\bw_3^1$, $\bw_3^2$, and $\bw_3^3$) for the octopole. To measure the alignment of the octopole with any given direction in the sky $\bOmega$, we can compute the following quantity:
\begin{equation}
\label{eq:S}
S(\bOmega) \equiv \frac{1}{3}\sum\limits_{i=1}^3\bOmega\cdot\bw_3^i.
\end{equation}
We are interested in two specific values of $S(\bOmega)$: its maximum, and $S(\bw_2)$. We denote them $S_\mathrm{O}$ and $S_\mathrm{QO}$ in the following. The first value measures the alignment of the three area vectors of the octopole and defines its averaged preferred direction, and the second value estimates the alignment between the quadrupole and octopole.

\begin{figure}
    \centering
    \includegraphics[width=0.99\linewidth]{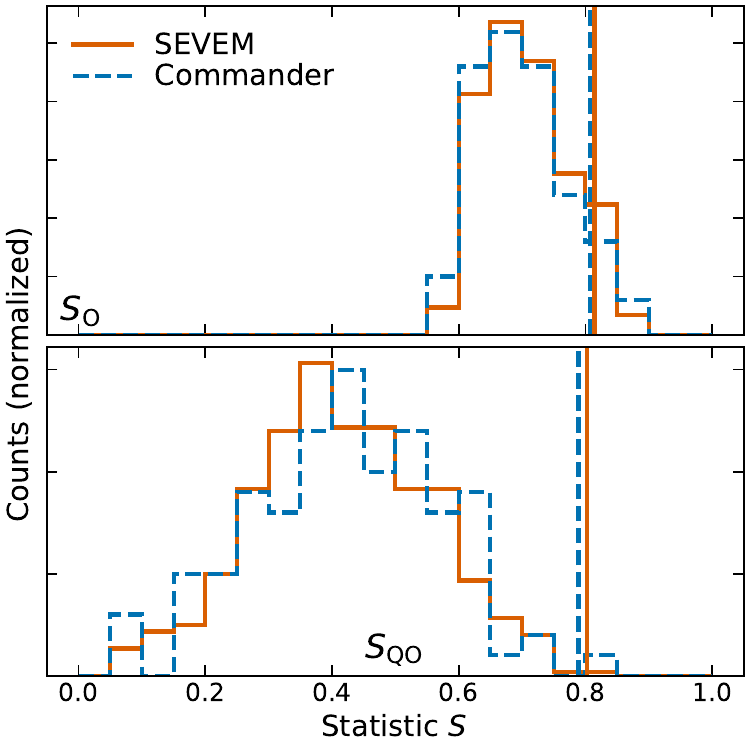}
    \caption{Quadrupole-octopole alignment in the cleaned \Planck\ PR4 CMB temperature maps. In the upper panel, we show the distribution of $S_\mathrm{O}$ (see eq.~\ref{eq:S}), indicating the planarity of the octopole is. Higher values are obtained for $59$ \sevem\ simulations (out of 600) and $9$ (out of 100) for \commander. In the lower panel, we show the distribution of $S_\mathrm{QO}$, measuring the alignment between quadrupole and octopole. Only one simulation in both sets has a higher $S_\mathrm{QO}$ value than the observed data. The solid orange and dashed blue lines correspond to the \sevem\ and \commander\ maps, respectively.}
    \label{fig:alignment-distribution}  
\end{figure}

In Fig.~\ref{fig:alignment-distribution} we show the distribution of $S_\mathrm{O}$ and $S_\mathrm{QO}$ determined from the \Planck\ PR4 simulated maps and their observed counterparts. Only $10\%$ of the simulations have a higher value of $S_\mathrm{O}$, which confirms that its three area vectors are relatively well aligned. This is the so-called planarity of the octopole. By itself, this does not constitute an anomaly, but it allows the high value of $S_\mathrm{QO}$, which is very close to $S_\mathrm{O}$. An alignment of the quadrupole and octopole like this is only found in one of the simulated CMB maps of each set. 

\begin{figure}
    \centering
    \includegraphics[width=0.99\linewidth]{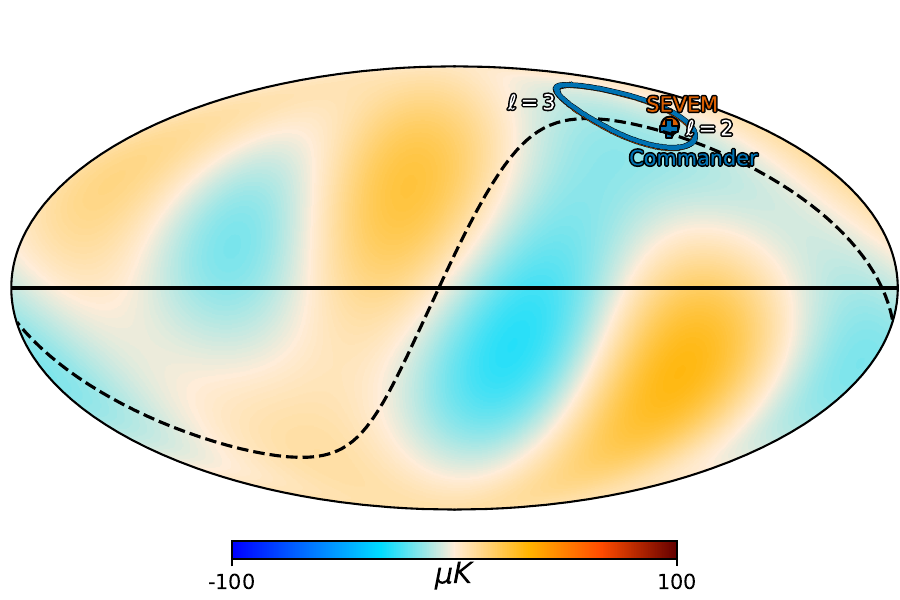}
    \caption{Quadrupole-octopole alignment in the cleaned \Planck\ PR4 CMB temperature maps. The preferred direction of the quadrupole is given by the orange circle and blue crosses for \sevem\ and \commander, respectively. The orange and blue contours, which are almost superimposed, correspond to the areas where $S(\Omega)$ (defined in Eq.~\ref{eq:S}) is within $3\%$ of its maximum (preferred direction of the octopole), for \sevem\ and \commander\ respectively. The background map is the sum of the \sevem\ CMB quadrupole and octopole. The solid and dashed black lines correspond to the galactic and ecliptic planes, respectively. We only show the preferred directions in the northern galactic hemisphere.}
    \label{fig:alignment-map}  
\end{figure}

In Fig.~\ref{fig:alignment-map} we show the preferred directions of the quadrupole and octopole as determined from the \Planck\ PR4 CMB maps after applying the common mask and inpainting the masked areas, as described in Sect.~\ref{sec:planck-data}. As expected from the measured values of $S_\mathrm{QO}$, the two favored axes point to the same area of the sky. The corresponding direction is close to the ecliptic plane and points toward the position of the Virgo cluster.

\subsection{Hemispherical asymmetry}
\label{sec:planck-asymmetry}

An efficient method for studying large-scale anisotropies of the CMB are position-dependent statistics. This consists of dividing the sky into a set of equal-size and uniformly distributed patches, and then computing patch by patch a chosen statistic for measuring its variation over the celestial sphere. Depending on the considered statistics as well as the patch properties (e.g., size and shape), different physical effects and scales can be probed.

For example, simply measuring the pixel variance in each patch is a very powerful probe of the hemispherical asymmetry \citep[known since WMAP; see e.g.][]{Eriksen:2003db}. This position-dependent variance (or local-variance) estimator was originally introduced by \citet{Akrami:2014eta}, and was applied to the different releases of \Planck\ data \citep[for a recent analysis of the PR4 data, see][]{Gimeno-Amo:2023jgv}. We extended the analysis to higher-order moments, namely the skewness and kurtosis. In Appendix~\ref{app:posdepcl} we also compute the position-dependent power spectrum.

Our patches were constructed from a very low resolution \healpix\ map ($\nside=8$) by setting one pixel to one and the rest to zero, and by upgrading its resolution to the same resolution as for the maps we analyzed (here $\nside=64$). Repetition of the process for each pixel of the low-resolution maps produced a set of $768$ patches that covered the full sky, without any overlapping areas. In Appendix~\ref{app:discs} we repeat our analysis considering disks with a radius of $4\degree$, which is the main patch choice of \citet{Planck:2019evm, Gimeno-Amo:2023jgv}. Both sets lead to the similar conclusions, with a higher statistical significance for the disks that were fine-tuned to study the hemispherical asymmetry.

We measured the temperature variance in patches of the \Planck\ PR4 CMB maps at low resolution (and the simulated CMB maps) after applying the common mask. Patches with more than $90\%$ of masked pixels were excluded following the analysis of \cite{Planck:2019evm}. For each patch of every map, we then subtracted the average value and divided by the variance computed from the corresponding set of simulations. The resulting maps are called position-dependent variances in the rest of the section. Position-dependent skewness and kurtosis maps were obtained following the same method.

\begin{figure}
    \centering
    \includegraphics[width=0.99\linewidth]{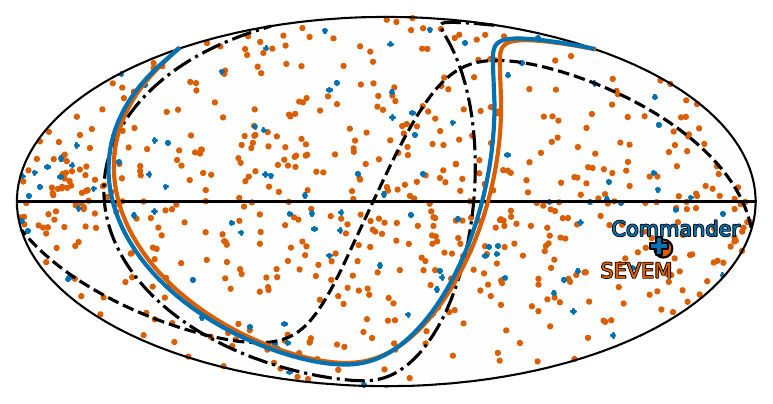}
    \caption{Preferred direction of the variance asymmetry in the \Planck\ PR4 CMB temperature maps. Dipoles are computed from the position-dependent variance maps computed using \healpix\ pixel patches. The data dipole directions are given by the large blue crosses and orange disks for \commander\ and \sevem, respectively, and the small markers are for the corresponding simulations. The blue and orange lines are the planes orthogonal to these observed dipoles. The solid, dashed and dash-dotted lines correspond to the galactic, ecliptic and supergalactic planes, respectively.}
    \label{fig:posdepvar-dipole}  
\end{figure}

To verify the presence of any hemispherical asymmetry, we fit a dipole to these position-dependent variances (and higher-order moments). Its amplitude $A_\mathrm{d}$ can confirm or refute the presence of an asymmetry in the maps, and its direction is orthogonal to the plane defining the two hemispheres. In Table~\ref{tab:posdepvar}, we compare the data dipole amplitude to the simulations, and we verify the well-known result that in the variance case, it is larger than for the vast majority of the simulated CMB maps at the $1$-$2\%$ level. No dipolar asymmetry is observed for the skewness and kurtosis. In Fig.~\ref{fig:posdepvar-dipole} we then show the corresponding directions on the sky of the position-dependent variance dipoles. The dipole directions of the different simulated CMB maps point all over the sky, which confirms that there is no preferred axis inherent to our analysis pipeline (e.g., masking choices). The observed asymmetry axis is relatively close to the supergalactic plane, where many CMB large-scale anomaliess of the local Universe can be observed (e.g., the Virgo, Coma, Centaurus, or Perseus clusters). In Sect.~\ref{sec:gaussian} we apply the same method to the local Universe simulations to determine whether its tSZ and kSZ contributions can explain the observed asymmetry.

\begingroup
\setlength{\tabcolsep}{6pt} 
\renewcommand{\arraystretch}{1.5} 
\begin{table}
    \caption{Dipole amplitude of position-dependent statistics}
    \begin{center}
      \begin{tabular}{|l|ccc|}
        \hline
         & Variance & Skewness & Kurtosis  \\
        \hline
         \sevem\ & $2.3$ & $50.8$ & $61.3$ \\
         \commander\ & $1$ & $55$ & $17$ \\
          \hline
        \end{tabular}
      \end{center}
    \tablefoot{Percentages of \Planck\ PR4 simulations with a larger dipole amplitude $A_\mathrm{d}$ of the position-dependent variance, skewness, and kurtosis than in observations, computed from $600$ simulations for \sevem, and $100$ for \commander, and using \healpix\ pixel patches. Only $1$ \commander\ simulation and $14$ \sevem\ simulations have a larger dipole amplitude for the variance.}
    \label{tab:posdepvar}
\end{table}
\endgroup

\par
\begin{figure*}
    \centering
    \includegraphics[width=0.99\linewidth]{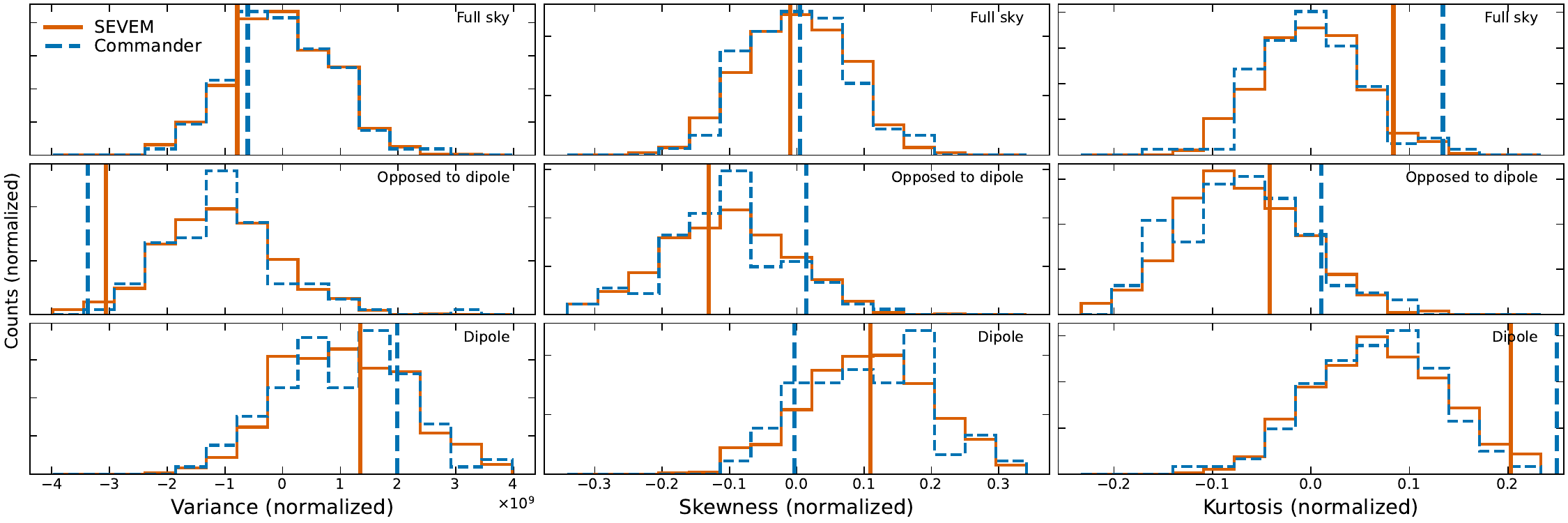}
    \caption{Hemispherical asymmetry in the \Planck\ PR4 CMB temperature data. From left to right, we show the position-dependent variance, skewness, and kurtosis averaged over the full sky (upper row), and the two hemispheres defined by the plane orthogonal to the position-dependent variance of each individual map (opposite and same directions as the dipole in the middle and lower rows, respectively). The vertical orange and blue lines correspond to the observed \sevem\ and \commander\ maps, respectively, and the histograms were obtained from the corresponding $600$ \sevem\ (orange), and $100$ \commander\ (blue) simulations. There is a lack of variance in the hemisphere opposed to the dipole, where $583$ (out of 600) \sevem\ and $99$ (out of 100) \commander\ simulations have higher values than the data.
    }
    \label{fig:posdep-distribution}  
\end{figure*}

\bigskip
Finally, we used the plane orthogonal to the measured dipole of each position-dependent statistics map in order to separate the map into two hemispheres, and we averaged it over these two halves of the sky. The results are shown in Fig.~\ref{fig:posdep-distribution}, where we compare the distribution of the different averaged quantities over the full sky, and the hemispheres in the dipole direction and opposite to it. In the direction opposite to the dipole in the observed data (close to the supergalactic northern hemisphere), the position-dependent variance is very low (among the $3\%$ lowest CMB simulated maps for the disks, as verified in Appendix~\ref{app:discs}), while the other hemisphere does not show any deviation. A splitting of the sky into two halves on either side of the ecliptic plane gives similar results, with a very low variance in one of the hemispheres. For the skewness, no such effect is seen, and this is also the case when splitting the sky using the ecliptic or supergalactic planes. In the \Planck\ map obtained from \commander\,, we observe a large position-dependent kurtosis, even in the full-sky case, and with a non-negligible difference with the \sevem\ results. We verified that the excess is localized in the supergalactic and ecliptic southern hemispheres, but no specific source can be easily identified. While there is a difference between the two hemispheres, no strong dipolar modulation was detected for the kurtosis, as reported in Table~\ref{tab:posdepvar}, and thus, we cannot directly relate this issue with the hemispherical asymmetry anomaly.

\section{Characterizing the tSZ and kSZ effects from the local Universe}
\label{sec:sz-analysis}

In this section, we further explore the statistical large-scale properties of the tSZ and kSZ effects from constrained simulations of the local Universe. In the first part, we focus on the power spectrum, and in the second part, we evaluate the impact of the secondary anisotropies induced by the local Universe on the different large-scale CMB anomalies studied in Sect.~\ref{sec:planck-analysis}. 

\subsection{SZ power spectra from the local Universe}
\label{sec:power-spectra}

We measured power spectra from the different maps shown in Fig.~\ref{fig:maps} (see also Table~\ref{tab:maps}). This included tSZ and kSZ maps for the two volumes of the local Universe ($\rmax=110,~350$ Mpc) and $\rmax=110$ Mpc maps, in which the position of the observer is slightly or strongly displaced in the \slow\ simulation. As expected, the tSZ signal was stronger by about one magnitude and dominated its counterpart kSZ effect at all scales.

For comparison purposes, we computed theoretical predictions for the tSZ power spectrum using the emulator described by \cite{Douspis:2021ing}, trained on the halo model developed by \cite{Salvati:2017rsn}. We also computed these predictions for the kSZ power spectrum using the numerical code \texttt{class\_sz}\footnote{\url{https://github.com/CLASS-SZ/class_sz}}\citep{Bolliet:2022pze},\footnote{\texttt{class\_sz} is based on the \texttt{class} code \url{http://class-code.net/} \citep{Blas:2011rf}} considering three different redshift ranges for the integration (two corresponded to the local Universe volumes, and the third to an integration up to $z=4$).

\begin{figure*}
    \centering
    \includegraphics[width=0.99\linewidth]{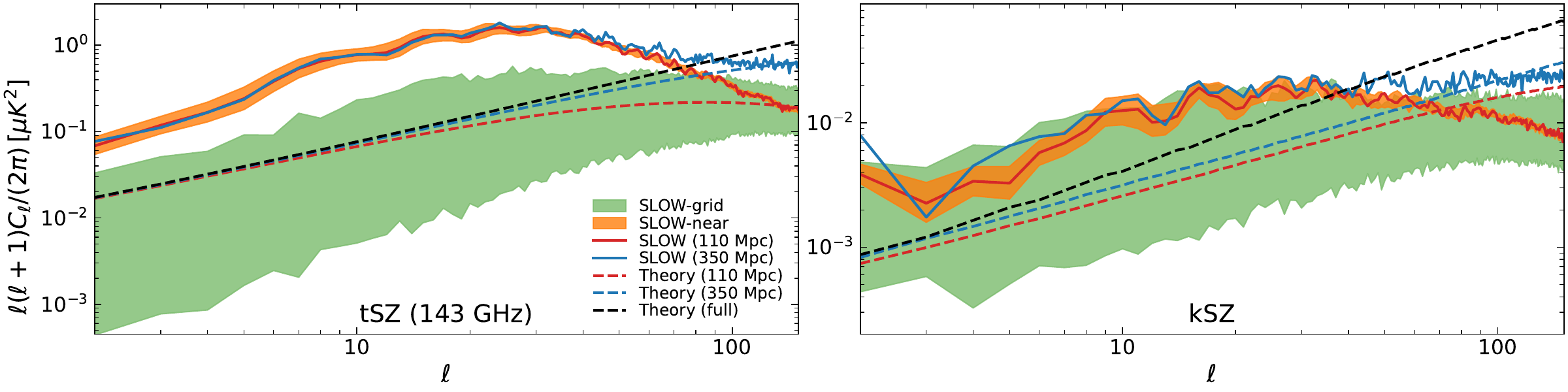}
    \caption{Local tSZ (left) and kSZ (right) power spectra compared to the average. The solid red and blue lines are power spectra measured from the maps shown in Fig.~\ref{fig:maps}, constructed by integrating electron gas properties through the \slow\ simulation from the position of the Milky Way to $\rmax=110$ and $\rmax=350$ Mpc, respectively. The colored areas are $90\%$ percentiles determined from the sets of maps where the center for the integration is displaced in the \slow\ simulation, slightly (by $1$ Mpc from the Milky Way) in orange (\slow-near), and strongly (more than $100$ Mpc between each center) in green (\slow-grid). The dashed lines show theoretical predictions computed in the halo model, where the red and blue lines are integrated up to $\rmax=110$ and $\rmax=350$ Mpc, and the black lines correspond to the full expected tSZ or kSZ signals. The tSZ power spectrum is shown at the frequency $143$ GHz.}
    \label{fig:power-spectra}  
\end{figure*}

In Fig.~\ref{fig:power-spectra} we compare these theoretical predictions to the power spectra from the simulated maps. We focus on the large scales ($\ell\leq150$), where the local tSZ and kSZ effect are expected to either dominate or to contribute significantly to the full SZ signal, as verified by theoretical calculations. This shows the agreement between these theoretical predictions and the range (defined as a $90\%$ percentile) that is spanned by the \slow-grid set, with a few exceptions that we discuss below.

For $\ell\leq70$, the tSZ power spectrum from the local Universe is well above the \slow-grid range, and it is larger by up to one magnitude than the theoretical prediction. The small uncertainties on the position of the Milky Way in the simulation that was obtained with the \slow-near set have a much weaker effect on this power spectrum, about $20\%$ at most. In addition, for $\ell<30$, the tSZ signal is almost entirely due to the very local Universe ($\rmax=110$ Mpc map). Similarly, the kSZ contribution of the local Universe is also at the high limit of the expected range, and again, the very local Universe represents a large part of the signal.

As we verified by applying different masks, the large amplitude of the local Universe tSZ power spectrum at low $\ell$ is related to the simulated replica of the Virgo cluster. We applied a disk mask with a radius of $10\degree$ centered on the Virgo cluster on the local Universe tSZ and kSZ maps (tests with smaller disks showed that they were not sufficient to remove the full contribution of Virgo). Similarly, we constructed one mask for each map of the \slow-grid and \slow-near sets to remove the contribution of the most prominent cluster in each map. This most prominent cluster was identified with the following procedure, which consisted of downgrading the tSZ maps to a resolution of $\nside=32$ and using the pixel with the highest value as the center of the disk mask. With this method, the Virgo cluster was always selected in the \slow-near simulations, as intended.

\begin{figure*}
    \centering
    \includegraphics[width=0.99\linewidth]{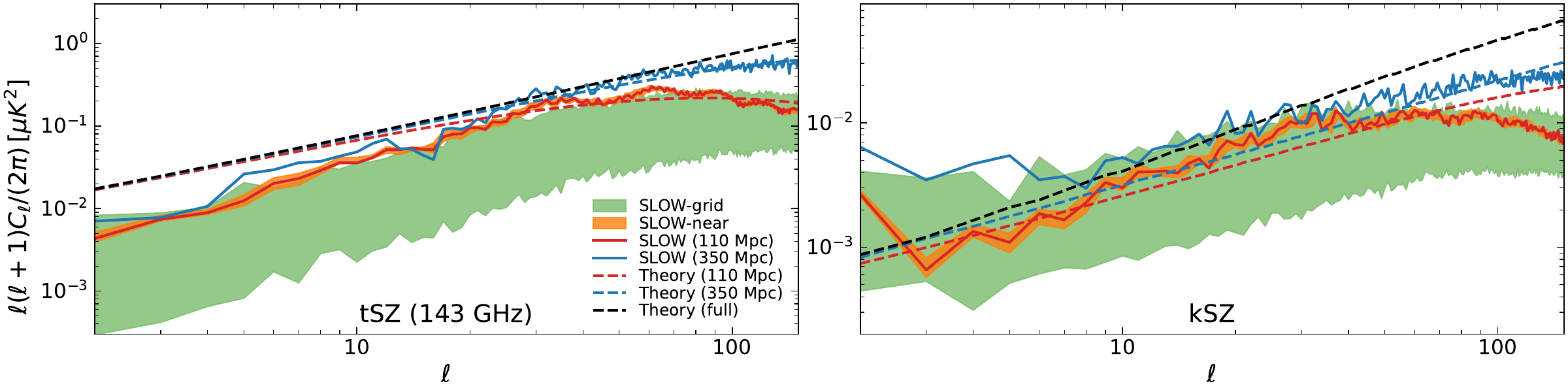}
    \caption{Same as Fig.~\ref{fig:power-spectra}, where the most prominent cluster in each simulation has been masked by a disk with a radius of $10\degree$. In the local Universe, this corresponds to the Virgo cluster.}
    \label{fig:power-spectra-virgo}  
\end{figure*}

The power spectra computed on these masked simulations of the tSZ and kSZ effects are shown in Fig.~\ref{fig:power-spectra-virgo}. When the contribution of the Virgo cluster is removed, the tSZ power spectrum at low $\ell$ decreases by more than one order of magnitude, bringing it in the range given by the \slow-grid set, but it is still in the high part. When Virgo is masked, the kSZ power spectrum is brought within the average on large scales. Moreover, masking the most prominent cluster has a weaker effect on most of the simulations than in the local Universe case. It decreases the tSZ signal by a factor $2$ on average. In the \slow-grid set of 27 simulations, only one simulation had a similar decrease after masking. This confirms the low probability of having a strong contribution from a cluster that is sufficiently close and as large as Virgo is.

\begin{figure*}
    \centering
    \includegraphics[width=0.99\linewidth]{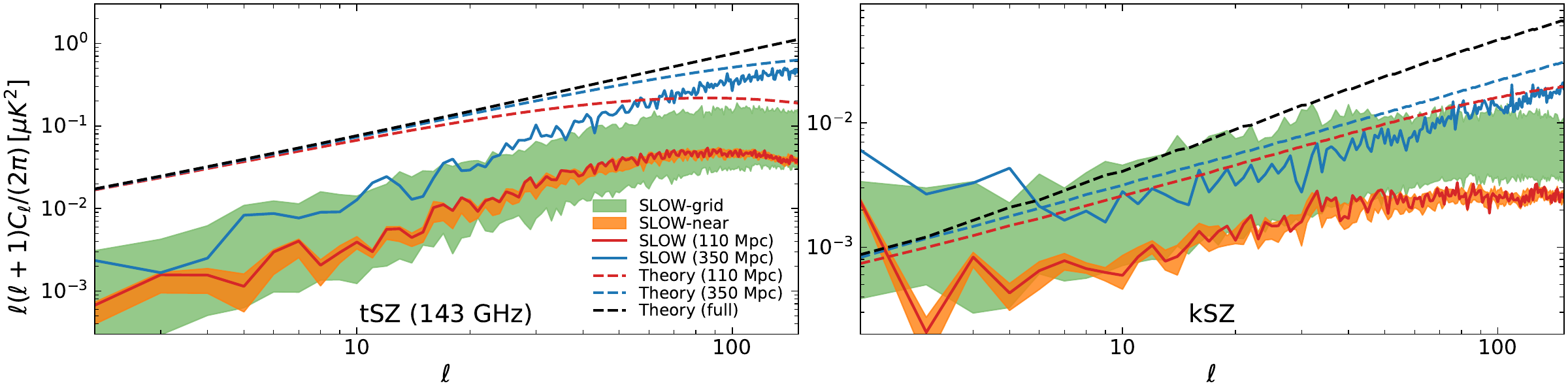}
    \caption{Same as Fig.~\ref{fig:power-spectra}, where the two most prominent clusters in each simulation have been masked by a disk with a radius of $10\degree$. In the local Universe, this corresponds to the Virgo and Centaurus clusters.}
    \label{fig:power-spectra-centaurus}  
\end{figure*}

We also individually masked several other identified clusters of the local Universe that were identified in \cite{Hernandez-Martinez:2024wxx}. The only cluster that contributed significantly to the power spectrum was Centaurus, which dominated the signal around $\ell \sim 50$. In Fig.~\ref{fig:power-spectra-centaurus} we verify that masking both Virgo and Centaurus indeed removes most of the \slow-110 tSZ and kSZ signals, and the difference with the \slow-350 power spectra finally becomes significant. In addition, the local Universe is now at the lower end of the \slow-grid range. This is consistent with the main conclusions of \cite{Sorce:2015yna, Dolag:2023xds}, who reported that the simulated local Universe is underdense up to $200$ Mpc, in agreement with observations \citep[see e.g.][]{Whitbourn:2013mwa}. The simulated equivalent of the Centaurus cluster of \slow\ has been shown by \cite{Hernandez-Martinez:2024wxx} to be almost five times more massive than expected, however. This means that its simulated SZ contribution is also larger.

\begin{figure*}
    \centering
    \includegraphics[width=0.99\linewidth]{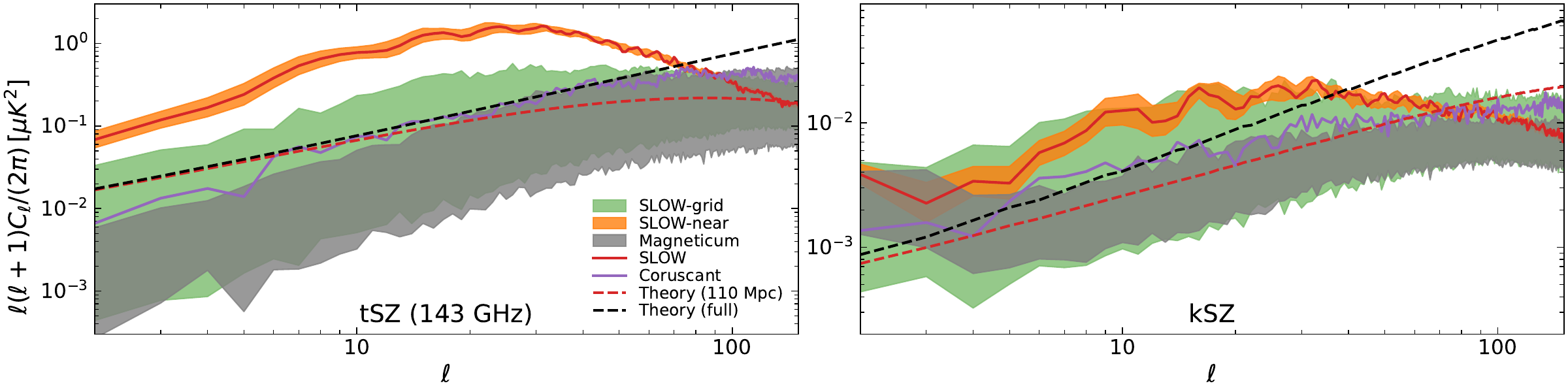}
    \caption{tSZ and kSZ power spectra up to $\rmax=110$ Mpc from several hydrodynamical simulations. As in Fig.~\ref{fig:power-spectra}, the solid red line and the orange and green areas are determined from the \slow\ simulation, and the dashed lines are the corresponding predictions from the halo model. The solid purple lines are estimated from local Universe maps determined from the \coru\ simulation, and the gray areas are $90\%$ percentiles from the \magn\ simulation.}
    \label{fig:power-spectra-all}  
\end{figure*}

\begin{figure*}
    \centering
    \includegraphics[width=0.99\linewidth]{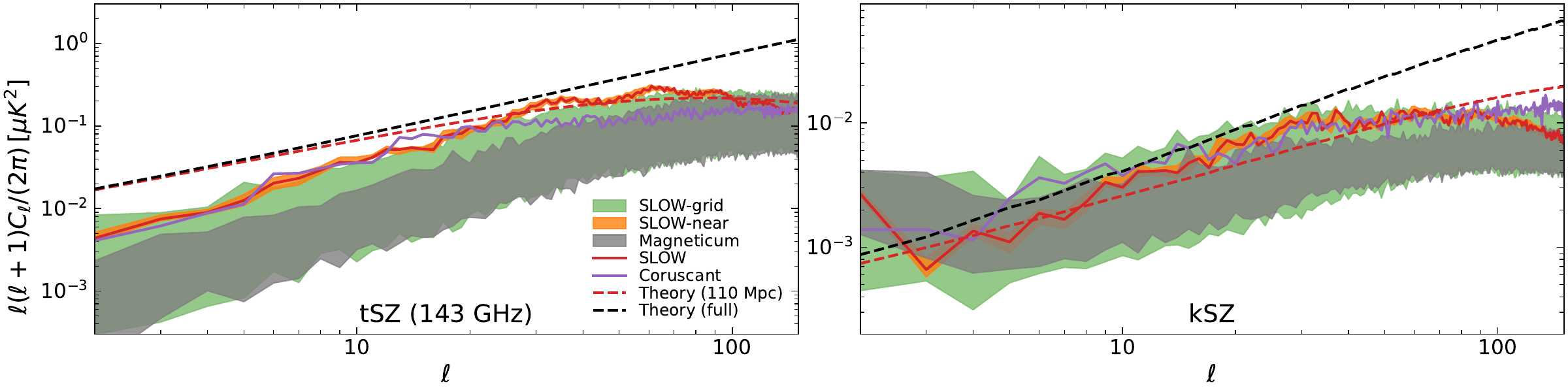}
    \caption{Same as Fig~\ref{fig:power-spectra-all}, where the most prominent cluster have been masked in each map.}
    \label{fig:power-spectra-all-masked}  
\end{figure*}

For further comparisons, we measured the power spectra from the maps derived from the constrained \coru\ and unconstrained \magn\ hydrodynamical simulations that we presented in Sects.~\ref{sec:coruscant} and \ref{sec:magneticum}, respectively. The corresponding results are shown in Figs.~\ref{fig:power-spectra-all} (full sky) and \ref{fig:power-spectra-all-masked} (with the most prominent cluster masked, e.g., Virgo in the case of the local Universe). Due to the different characteristics of each simulation (cosmological parameters, resolution, box sizes), no perfect agreement is expected. To take the dependence on cosmological parameters into account, we rescaled the tSZ power spectra of the \coru\ and \magn\ maps using $C_\ell^\mathrm{tSZ} \propto \sigma_8^3 \Omega_m^3$ \citep{Hill:2013baa}. Similarly, the \coru\ and \magn\ kSZ power spectra were rescaled using the factor between the theoretical predictions obtained for the different cosmologies.

One important difference between the two generations of constrained simulations (\slow\ and \coru) is the impact of the Virgo cluster on large scales. In \coru, no excess signal is observed on these scales, which means that it is lower by one order of magnitude than \slow. However, after masking Virgo, the tSZ and kSZ power spectra of both maps match extremely well up to $\ell \sim 20$ (tSZ) and $\ell \sim 100$ (kSZ), before other effects such as the resolution of the box play a role. For the kSZ power spectrum, removing the Virgo contribution also has a much stronger effect in \slow. 

Furthermore, the \slow-grid set of maps spans a wider range of values than its equivalent set from \magn\ on large scales ($\ell \leq 30$), which indicates that a cluster like Virgo is even more unlikely in the \magn\ set (close and large enough to entirely dominate the signal). One possible explanation is that, as reported in \citet{Sorce:2015yna, Dolag:2023xds}, the local Universe contains a significant overdensity of very massive clusters (masses above $10^{15}~M_\odot$) in a large volume around our position (up to $\rmax=200$ Mpc) with respect to the \magn\ simulation (although not exactly the same as here). This might make a massive cluster in the neighborhood of the different centers used in the \slow-grid set more likely.

In conclusion, this highlights two interesting properties of the local tSZ and kSZ power spectra on large scales compared to other volumes of the same size selected randomly. The local power spectra are dominated by a very low number of clusters, mainly Virgo and Centaurus. Despite this, the low-$\ell$ tSZ power spectrum is larger than in the majority of the other simulated maps (around $95\%$), mainly because the Virgo cluster is so close to us. Its signal expands on a large area of the sky, as a disk with a radius of $10\degree$ is necessary to mask it fully, which is much wider than the disk radius of $2\degree$ of the \Planck\ common mask, for example. Therefore, we evaluate in the following section whether this larger-than-expected local SZ signal can significantly modify the statistical properties of the CMB on very large angular scales.

\subsection{Local tSZ and kSZ effects and CMB large-scale anomalies}
\label{sec:gaussian}

We used the simulated maps of the tSZ and kSZ effects from the local Universe, shown in Fig.~\ref{fig:maps}, to investigate their impact on the largest angular scales of the CMB. On these scales, several so-called anomalies (see Sect.~\ref{sec:planck-analysis}) have been observed, with indications of a correlation with the local Universe (e.g., alignments of low multipoles near the Virgo direction, or an asymmetry axis close to the supergalactic plane).

For this purpose, we produced CMB realizations including the local tSZ and kSZ effects. We started by generating $10000$ CMB Gaussian realizations from the \Planck\ best-fit theoretical power spectrum computed with \texttt{CAMB}, with a resolution of $\nside=64$ and a Gaussian beam with an FWHM of $160'$, which is the same as for the \Planck\ maps described in Sect.~\ref{sec:planck-data}. To each of these Gaussian CMB simulations, we added the tSZ and kSZ signal from the local Universe obtained from the constrained simulations at a chosen frequency, after degrading the map resolution to $\nside=64$ and applying the $160'$ Gaussian beam (again, the same as for the \Planck\ maps). We masked the galactic plane, and in some cases, Virgo (with a disk radius of $10\degree$), keeping almost $70\%$ of the sky for the analysis. The dipole of the resulting maps was then removed, which minimized the impact of any potential systematics at the dipole level in the initial conditions of the reconstruction procedure. Finally, we computed the different estimators of CMB large-anomalies described in Sect.~\ref{sec:planck-analysis} in the Gaussian CMB simulations and in their CMB + SZ counterparts to compare the differences.

We report our main results in Figs.~\ref{fig:sz_and_anomalies-correlation}, \ref{fig:sz_and_anomalies-alignment}, and \ref{fig:sz_and_anomalies-hpa} and focus on the lack of correlation, quadrupole-octopole alignment, and hemispherical asymmetry anomalies, respectively. We used the $\rmax=350$ Mpc maps as our baseline, with a frequency of $143$ GHz for the tSZ signal. {At this frequency, the tSZ signal entirely dominates its kSZ counterpart, as shown in Sect.~\ref{sec:power-spectra}. Therefore, the latter is not expected to play a significant role at very large scales. Still, at galaxy cluster scales (a few to a few dozen arcminutes), kSZ can impact the total signal of individual clusters depending on the direction of their bulk velocity. Further tests with other frequencies and in particular, at $217$ GHz, where the tSZ signal is suppressed, led to very similar conclusions.  We also verified that the lower $\rmax$ maps and a full-sky analysis did not impact our results significantly. 

The main conclusion was the same for the three anomalies. The SZ effect from the local Universe affects the different quantities by a few percent at most, which is far below the level of the anomalies. For example, the observed $S_{1/2}$ is lower than half of the lowest value of all the simulated \Planck\ CMB maps (which itself is a few times lower than the average). Similarly, moving the observed $S_{QO}$ by $2\%$ would still leave it in the tail of the distribution of alignments in the simulated \Planck\ CMB maps. Moreover, there is no visible preferred sign, meaning that its effect on the distribution of the different statistics is even much weaker on average. 

When we focus on the simulations that are "anomalous" (i.e., the different estimators are in the tail of the expected distribution, at a level similar to the \Planck\ estimators), the impact of the SZ effect from the local Universe is even smaller. The effect is slightly stronger at most when Virgo is not masked at all, meaning that despite its domination of the SZ signal at low multipoles, its effect on CMB large-scale anomalies is only very weak. 

\begin{figure}
    \centering
    \includegraphics[width=0.99\linewidth]{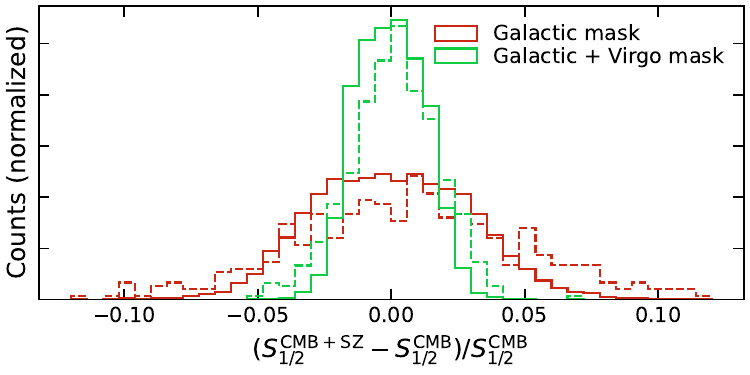}
    \caption{Impact of the local Universe SZ signal on the lack of correlation at angular scales $\theta>60\degree$. We measure the difference of $S_{1/2}$ (see Eq.~\ref{eq:S1/2}) between Gaussian CMB realizations with or without the local tSZ (at $143$ GHz) and kSZ effects, when the galactic plane alone was masked (in red), and when the galactic plane and Virgo were masked (in green). The distributions are shown for both the full set (solid lines) and for the $5\%$ (dashed) simulations where the lack of correlation is strongest (smallest $S_{1/2}$).
    }
    \label{fig:sz_and_anomalies-correlation}  
\end{figure}

\begin{figure}
    \centering
    \includegraphics[width=0.99\linewidth]{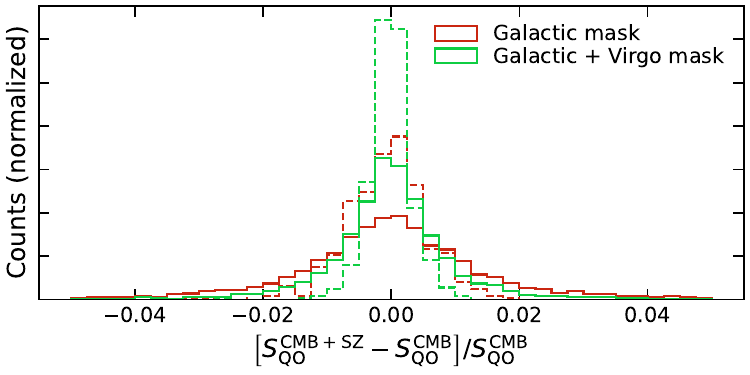}
    \caption{Impact of the local Universe SZ signal on the quadrupole-octopole alignment. We measure the difference of $S_\mathrm{QO}$ (see Eq.~\ref{eq:S}) between Gaussian CMB realizations with or without the local tSZ (at $143$ GHz) and kSZ effects, when the galactic plane alone was masked (in red), and when the galactic plane and Virgo were masked (in green). The distributions are shown for both the full set (solid lines) and the $5\%$ (dashed) simulations where the alignment is strongest (largest $S_\mathrm{QO}$). 
}
    \label{fig:sz_and_anomalies-alignment}  
\end{figure}

\begin{figure}
    \centering
    \includegraphics[width=0.99\linewidth]{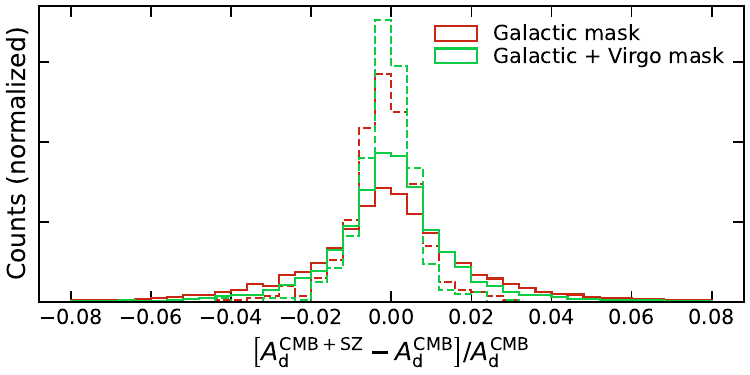}
    \caption{Impact of the local Universe SZ signal on the hemispherical asymmetry. We measure the difference of $A_\mathrm{d}$ (amplitude of the position-dependent variance dipole; see Sect.~\ref{sec:planck-asymmetry} for details) between Gaussian CMB realizations with or without the local tSZ (at $143$ GHz) and kSZ effects, when the galactic plane alone was masked (in red), and when the galactic plane and Virgo were masked (in green). The distributions are shown for both the full set (solid lines) and the $5\%$ (dashed) simulations where the hemispherical asymmetry is strongest (largest $A_\mathrm{d}$).
    }
    \label{fig:sz_and_anomalies-hpa}  
\end{figure}

\section{Conclusion}
\label{sec:conclusion}

We have explored the impact of the local Universe through the tSZ and kSZ effects on the CMB temperature anisotropies at large scales. Our objective was twofold: to characterize the large-scale properties of the CMB temperature fluctuations with a new analysis of the large-scale anomalies of the latest \Planck\ data, and to verify their correlation with the local tSZ and kSZ signals as obtained from advanced constrained hydrodynamical cosmological simulations.

Using the constrained hydrodynamical simulation \slow\ to reproduce a large volume of the nearby LSS, we constructed high-resolution maps of the local tSZ and kSZ effects. We showed that they are the dominant contributions of the overall expected tSZ and kSZ effects on large angular scales by comparison to theoretical predictions, which leaves the anisotropic features correlated to the local Universe in the large-scale fluctuations of the CMB. We characterized the local tSZ and kSZ signals mainly by studying their power spectrum, highlighting their strong dependence on a few specific structures in the maps. First tests with higher-order statistics such as the bispectrum, which are beyond the scope of this work and will be discussed in future works, led to similar conclusions. In particular, the Virgo cluster causes most of the large-scale signal. Moreover, this signal is stronger than in $\sim 95\%$ of the other random patches of the Universe that we tested, regardless of whether we moved the observer position in the \slow\ simulation or used the unconstrained \magn\ simulation. The localization and the amplitude of the local Universe tSZ and kSZ signals, which is higher than expected, mean that further investigation of their impact on the large-scale CMB temperature fluctuations is required, for example, based on the methods developed to study the CMB large-scale anomalies.

We have analyzed several of the CMB temperature large-scale anomalies in the latest \Planck\ temperature data (PR4), the lack of correlation using the quantity $S_{1/2}$, the quadrupole-octopole alignment using multipole vectors and the hemispherical asymmetry with a local variance approach. Our analysis was conducted on cleaned CMB maps from both \commander\ and \sevem and led to similar conclusions as other analyses that were conducted on previous datasets \citep[see e.g.,][for PR3 results]{Planck:2019evm}, or \cite[][on PR4]{Gimeno-Amo:2023jgv, Billi:2023liq}. These unexpected deviations from isotropy are observed at a level found in typically less than $1\%$ of the \Planck\ CMB simulated maps, with some properties, such as a preferred direction of the quadrupole and octopole in the Virgo region and an asymmetry axis that is very close to the supergalactic plane. This indicates that the local LSS is a possible explanation.

We have used the simulated maps of the tSZ and kSZ effects from the local Universe, together with Gaussian CMB realizations based on the \Planck\ best-fit cosmological parameters, to estimate the impact of these secondary anisotropies on the largest angular scales of the CMB fluctuations. Our analysis was based on the same estimators as the analysis of large-scale anomalies in the \Planck\ data, and we  showed that the local tSZ and kSZ effects cannot explain the detected deviations from isotropy.

\begin{acknowledgements} The authors thank the referee for their comments and suggestions.
We thank Adélie Gorce, Andrea Ravenni and Saleem Zaroubi for useful inputs and discussions.
This work was supported by the grant agreements ANR-21-CE31-0019 / 490702358 from the French Agence Nationale de la Recherche / DFG for the LOCALIZATION project. NA acknowledges funding from the ByoPiC project from the European Research Council (ERC) under the European Union’s Horizon 2020 research and innovation program grant number ERC-2015-AdG 695561. KD acknowledges support by the COMPLEX project from the European Research Council (ERC) under the European Union’s Horizon 2020 research and innovation program grant agreement ERC-2019-AdG 882679. The authors acknowledge the Gauss Centre for Supercomputing e.V. (\url{www.gauss-centre.eu}) for funding this project by providing computing time on the GCS Supercomputer SuperMUC-NG at Leibniz Supercomputing Centre (\url{www.lrz.de}) under the projects pr86re, pr83li and pn68na. This research made use of observations obtained with \Planck\ (\url{http://www.esa.int/Planck}), an ESA science mission with instruments and contributions directly funded by ESA Member States, NASA, and Canada. The authors acknowledge the use of the healpy and \healpix\ packages \citep{Zonca2019, 2005ApJ...622..759G}.
\end{acknowledgements}

\bibliographystyle{aa}
\bibliography{biblio}

\begin{appendix}

\section{The \Planck\ position-dependent power spectrum}
\label{app:posdepcl}

In addition to the position-dependent variance, skewness and kurtosis analyses presented in Sect.~\ref{sec:planck-asymmetry}, we consider the position-dependent power spectrum. When used up to small scales where noise becomes dominant, it has been shown to be an efficient ingredient to study several signatures of non-Gaussianity, like the primordial local template or the correlations between the ISW and lensing effects \citep[see][and references therein]{Jung:2020zne}. Here, in the spirit of this paper, we focus only on large angular scales ($\ell \leq 50$), and the rest will be included in a future work. 

For this analysis, we use a set of patches based on Mexican needlets \citep[see][for details]{Scodeller:2010mp}, which have the advantage of having simple expressions in harmonic space, but are not exactly localized in real space, unlike the disks and pixels. A needlet patch map centered at the position $\bm{\Omega}_\mathrm{c}$ is obtained using
\begin{equation}
    \label{eq:mexican-needlet}
     M^\mathrm{patch}_{p, B, j}(\bm{\Omega};\bm{\Omega}_\mathrm{c}) 
     = \sum_\ell \frac{2\ell+1}{4\pi} \left[\frac{\ell(\ell+1)}{B^{2j}}\right]^p e^{\frac{-\ell(\ell+1)}{B^{2j}}} P_\ell(\bm{\Omega} \cdot \bm{\Omega}_\mathrm{c}),
\end{equation}
where $P_\ell$ is a Legendre polynomial and, $p$, $B$ and $j$ are the needlet parameters. Here we use the values $p=1$, $b=4$ and $j=1$, for which only terms with $\ell \leq 10$ are sufficient to compute the sum, and $192$ patches are enough to cover uniformly the full sky. In Fig.~\ref{fig:patches}, we show an example of such needlet patch, along with the pixels and disks used in Sects.~\ref{sec:planck-asymmetry} and \ref{app:discs}, respectively.

\begin{figure*}
    \centering
    \includegraphics[width=0.99\linewidth]{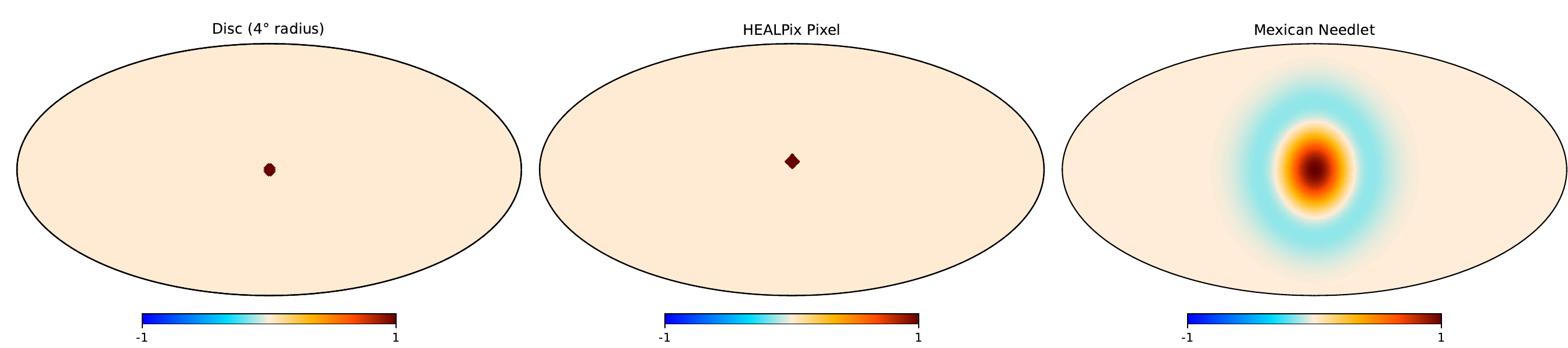}
    \caption{Maps of the disk, \healpix\ pixel and Mexican needlet patches used in this work to evaluate position-dependent statistics. The three complete sets include $3072$ $4\degree$ radius disks, $768$ pixels (corresponding to the pixels of a $\nside=8$ \healpix\ map) and $192$ Mexican needlets ($p=1$, $b=4$ and $j=1$). Note that all patches have been set to $1$ at their center in this figure.}
    \label{fig:patches}  
\end{figure*}

As in Sect.~\ref{sec:power-spectra}, we measured pseudo-$C_\ell$ up to $\lmax=50$, using \texttt{NaMaster} in each of the $192$ needlet patches defined above. For each multipole of the position-dependent power spectrum, we measured a dipole (following the same pipeline as for the position-dependent variance). All the corresponding directions are shown in Fig.~\ref{fig:posdepcl-dipole}. An important result is that all the dipoles are localized in less than half the sky, as they point in both the southern ecliptic and supergalactic hemispheres (at the exception of $\ell=47$, slightly over the ecliptic plane), and the angular difference between close multipoles is almost always small, indicating some correlation between them. However, opposed to the variance case, the amplitude of the data dipoles is in the expected range obtained from the simulations, meaning this estimator is not sensitive to the hemispherical asymmetry.

\begin{figure}
    \centering
    \includegraphics[width=0.99\linewidth]{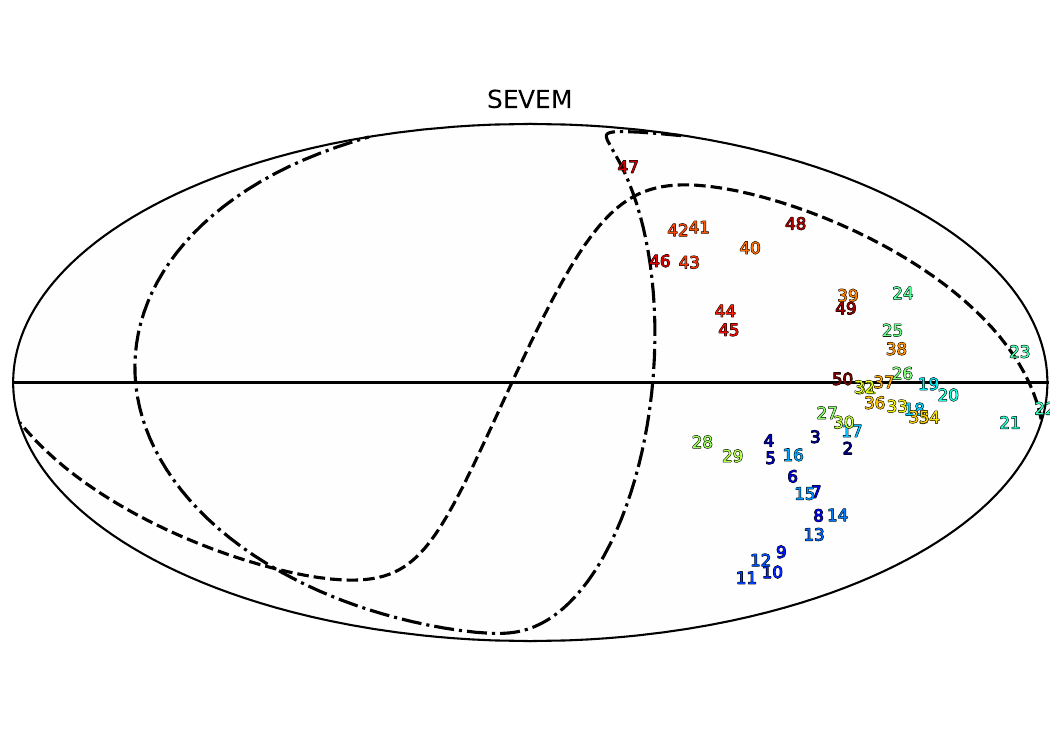}
    \caption{Dipole directions of the position-dependent power spectrum in the \Planck\ PR4 \sevem\ map (\commander\ results being very similar), for each multipole up to $\ell=50$. The black solid, dashed and dash-dotted lines correspond to the galactic, ecliptic and supergalactic planes, respectively.}
    \label{fig:posdepcl-dipole}  
\end{figure}

In Fig.~\ref{fig:posdepcl-distribution} we compare the observed position-dependent power spectrum to the distribution determined from the corresponding simulations. On the largest scales ($\ell=2, 3$ and $4$), the observed value is close to the $5\%$ lowest simulations, which is related to the lack-of-power/lack-of-correlation anomaly. When repeating the same analysis on half the sky, using the dipole directions determined above, we find that the hemisphere opposed to the dipole has a low power up to $\ell=7$, and is among the $1$-$2\%$ range on the largest scales. However, the other hemisphere also has a relatively low power at low multipoles ($20\%$ lowest simulations).

\begin{figure}
    \centering
    \includegraphics[width=0.99\linewidth]{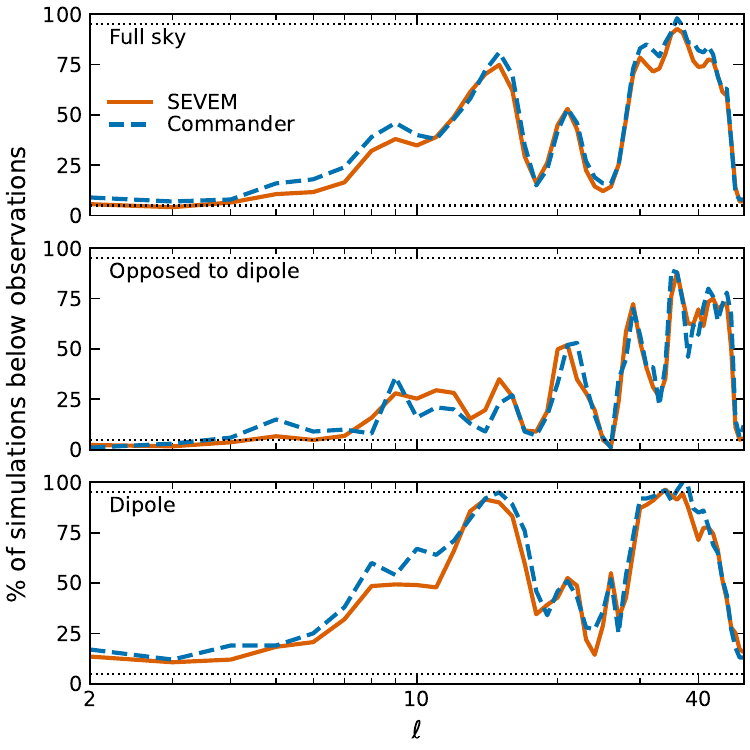}
    \caption{Comparison between the position-dependent power spectrum in the \Planck\ PR4 CMB temperature data and simulated maps. We average the position-dependent power spectrum over the full sky (upper panel), and the two hemispheres defined by the dipole of each multipole (opposed to dipole in the middle panel and dipole direction otherwise), and compare the observed value to the equivalent from the simulations. We show the percentage of simulations having a lower value than the data. Horizontal black dotted lines indicate the $5\%$ and $95\%$ thresholds.}
    \label{fig:posdepcl-distribution}  
\end{figure}

\section{The \Planck\ position-dependent variance with disks}
\label{app:discs}

Here we report the main results of our analysis of the position-dependent variance of the \Planck\ PR4 CMB temperature maps, where we use disks of $4\degree$ radius as patches.

The conclusions from Figs.~\ref{fig:posdep-dipoles-discs} and \ref{fig:posdep-distribution-discs} are very similar to the ones reported in Sect.~\ref{sec:planck-asymmetry} with the \healpix\ pixel patches. The large dipole of the observed position-dependent variance confirms the presence of a strong hemispherical asymmetry, maximized when separating the sky close to the supergalactic plane, with a significant lack of variance in the northern half. Note that with the disk patches, this anomaly is stronger than in any of the simulations. For further analyses of the hemispherical asymmetry in the PR4 data, including polarization, with the same disk patches, we refer the reader to \citet{Gimeno-Amo:2023jgv}. 

\begin{figure}
    \centering
    \includegraphics[width=0.99\linewidth]{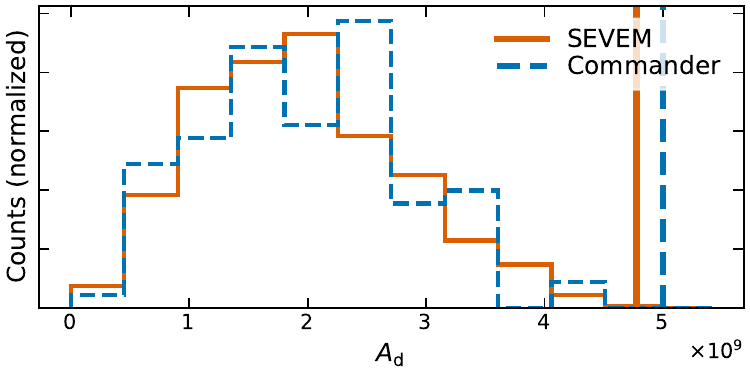}
    \caption{
    Distribution of the amplitude of the position-dependent variance dipole (using disks of $4\degree$ radius) in the \Planck\ PR4 CMB temperature data illustrating the hemispherical asymmetry anomaly. The orange and blue vertical lines correspond to the observed \sevem\ and \commander\ maps, respectively, and the histograms are obtained from the corresponding $600$ \sevem\ (orange), and $100$ \commander\ (blue) simulations. None of the simulated CMB maps have a large dipole amplitude value.
    }
    \label{fig:posdep-dipoles-discs}  
\end{figure}

\begin{figure}
    \centering
    \includegraphics[width=0.99\linewidth]{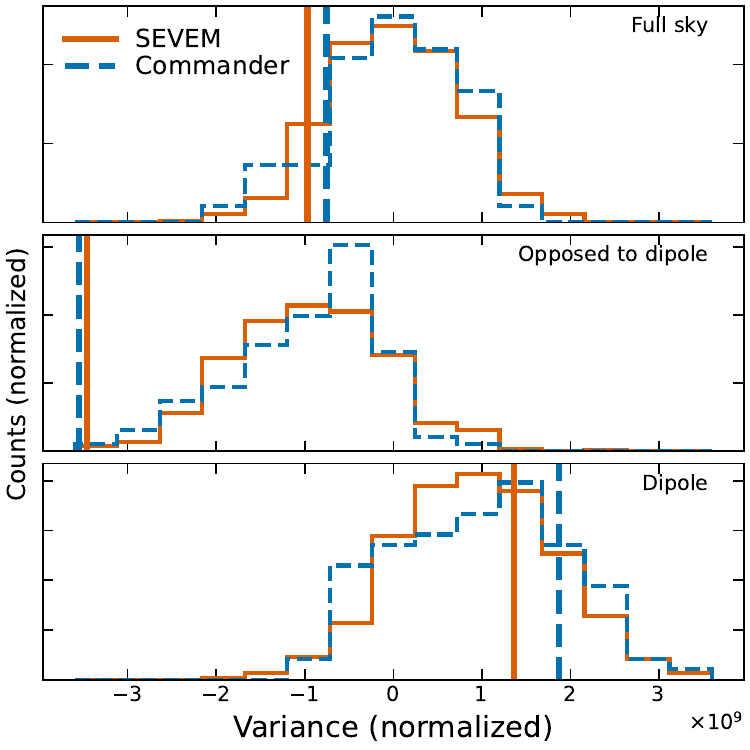}
    \caption{
    Same as Fig.~\ref{fig:posdep-distribution} but for the position-dependent variance computed with disks of $4\degree$ radius as patches. None of the simulated CMB maps have a lower averaged position-dependent variance in the direction opposed to their respective dipoles than the observations (close to the northern supergalactic hemisphere).
    }
    \label{fig:posdep-distribution-discs}  
\end{figure}

\end{appendix}

\end{document}